\documentstyle[11pt,psfig]{article}

\def\threeR{{}^{(3)}\kern-1pt R}
\def\three{{}^{(3)}\kern-1pt}

\def\fraction#1#2{{\textstyle{#1\over#2}}} \def\fr{\fraction}
 
\def\vol#1{{\bf #1}}
\def\journalfont{\it}         
\def\jou#1{{\journalfont #1\ }}

\def\cmp{\jou{  Commun.\ Math.\ Phys.}}

\def\grg{\jou{  Gen.\ Relativ.\ Grav.}}

\def\pr{\jou{   Phys.\ Rev.}}
\def\prep{\jou{ Phys.\ Rep.}}

\def\cgr{0420}            
\def\rag{9530S}               
\def\cosmology{9880H}     
\def\book#1{{\it #1\/}}   \def\pub(#1){(#1)}
\newcommand{\support}[1]{\thanks{#1}\ }

\newcommand{\be}{\begin{equation}}
\newcommand{\ee}{\end{equation}}

\catcode`@=11

\newdimen\jot \jot=3pt
\newskip\z@skip \z@skip=0pt plus0pt minus0pt
\newdimen\z@ \z@=0pt 
\dimendef\dimen@=0

\def\m@th{\mathsurround=\z@}

\def\ialign{\everycr{}\tabskip\z@skip\halign} 

\def\openup{\afterassignment\@penup\dimen@=}
\def\@penup{\advance\lineskip\dimen@
  \advance\baselineskip\dimen@
  \advance\lineskiplimit\dimen@}

\def\eqalign#1{{
\baselineskip=20pt \lineskip=10pt \lineskiplimit=4pt
\null\,\vcenter{\openup\jot\m@th
  \ialign{\strut\hfil$\displaystyle{##}$&$\displaystyle{{}##}$\hfil
      \crcr#1\crcr}}\,} }

\def\meqalign#1{\null\,\vcenter{\openup\jot\m@th
  \ialign{\strut\hfil$\displaystyle{##}$&&$\displaystyle{{}##}$\hfil
      \crcr#1\crcr}}\,}
\catcode`@=12   

\def\Hfunc{1+\left( \gamma-1 \right)v^2}
\def\Sp{\Sigma_+}
\def\A1{A_1}
\def\f{{\rm fin.}}
\def\eqp#1{
  \vskip 3 mm
  \noindent
  {\it The equilibrium point\/} #1:
  }

  \def\Strut{\vrule width 0pt height 13pt depth 6pt{}}

\def\elim{\rm eliminated}

\begin{document}
\bibliographystyle{unsrt}  


\title{Spatially self-similar locally rotationally
       symmetric perfect fluid models}

     \author{Ulf Nilsson
       {\support{e-mail: ulfn@vanosf.physto.se}}\\
       {\it Department of Physics, Stockholm University,}\\
       {\it Box 6730, S-113 85 Stockholm, Sweden} \\
        and \\
       Claes Uggla
       \support{Supported by the Swedish Natural Science Research
         Council}
       \support{e-mail: uggla@vanosf.physto.se}\\
       {\it Department of Physics, Stockholm University,}\\
       {\it Box 6730, S-113 85 Stockholm, Sweden} \\
       and \\
       {\it Department of Physics, Lule\aa\ University of
       Technology}\\
       {\it S--951 87 Lule\aa, Sweden} }
\maketitle

\begin{abstract}
Einstein's field equations for spatially self-similar locally rotationally
symmetric perfect fluid models are investigated. The field equations
are rewritten as a first order system of autonomous
ordinary differential equations.
Dimensionless variables are chosen in such a way
that the number of equations in the coupled system of differential equations
is reduced as far as
possible. The system is subsequently analyzed qualitatively for some of the
models. 
The nature of the singularities occurring in the models is discussed.
\end{abstract}

PACS numbers: \cgr, \rag, \cosmology.

\vspace{3.0cm}

Short title: Spatially self-similar perfect fluid models

\newpage



\section{Introduction}

Self-similar models, with a group acting transitively on 3-dimensional
hypersurfaces, has been discussed in the literature for several decades.
Within this family of models one finds a number of different interesting
physical phenomena; chock waves and violation of cosmic censorship being
perhaps the most prominent ones.
Self-similar models are also interesting because they constitute
asymptotic states for more general models and that they thus act as building
blocks when it comes to understanding wider classes of models.
They also generalize the spatially homogeneous (SH) models which admit two
spacelike commuting Killing vectors, and
in this context they can be seen as the first step
towards the construction of inhomogeneous cosmological models.
They might also shed light on the generality of properties
found in the SH subclass of models.
For example, in this article we discuss if whimper
singularities, 
found in some of the SH models, exist in the self-similar case.

One finds two approaches toward the self-similar perfect fluid models.
One is the ``fluid adapted'' approach in which one choses a timelike
coordinate along the fluid lines (see e.g., \cite{ctau}--\cite{ori}).
The second is the ``homothetic''
approach in which one chooses a coordinate along the orbits of the
homothetic Killing vector (see e.g., \cite{ear}--\cite{wu}).
In general, the symmetry surface changes causality.

One often considers models that are diagonalizable. In this case
the homothetic approach, taken together with the diagonal gauge requirement,
has the drawback that
one has to cover the spacetime with several coordinate patches.
On the other hand, the homothetic approach reveals that there is a
considerable structural similarity between the field equations for the
self-similar models and
the hypersurface-homogeneous models (e.g., SH models).
There is a considerable literature about on how one deals with
hypersurface-homogeneous models (much larger than the one on self-similar
models). Thus the homothetic approach makes it possible to transfer ideas from
the hypersurface-homogeneous arena to the self-similar one.
Because of this advantage we will use the homothetic approach.
However,
note that results obtained in this picture can be transferred to the
fluid adapted formulation by a coordinate transformation
and vice versa.

In this article we will focus on the ``spatial part'' of
self-similar models exhibiting a locally rotationally symmetric (LRS)
isometry group (in addition to the self-similar symmetry).
The line elements for the spatially-self-similar (SSS) LRS models have been
given by Wu \cite{wu}. Collectively they can be written as
\be
   d{\tilde s}^2 = e^{-2fx} ds^2 = e^{-2fx}\left[
   -dt^2 + D_1(t)^2 dx^2 + D_2(t)^2 e^{-2ax}\left( dy^2+
   k^{-1}\sin{(\sqrt{k}y)}dz^2 \right)\right]
\ee
where $f,a,k$ are parameters describing the symmetry groups of the various
models. Canonical values for these parameters are given in Table 1. Note that
$ak = 0$ and that, for the sake of brevity, we have denoted the LRS type
$_f$I ($_f$V) models with $_f$I ($_f$V) even though they also contain a
type $_f$VII$_0$ ($_f$VII$_h$) group
(see \cite{wu}). The SSS models with spherical symmetry are denoted by
$^*$KS where KS stands for Kantowski-Sachs in analogy with the
corresponding SH case.

\begin{table}[ht]
\begin{center}
  \begin{tabular}{|c|c|c|c|c||c|c|c|c|}  \hline 
    \Strut$\mbox{}$ & \multicolumn{4}{c||}{SSS} &
    \multicolumn{4}{c|}{SH} \\ \hline \Strut & $_f$V
    &$\mbox{}_1^*$III&$\mbox{}^*$KS &$\mbox{}_1$I & V & III & KS & I
    \\ \hline
    \Strut$f$ & $f$ & $-1$ & $-1$ & $-1$ & 0 & 0 & 0 & 0 \\ \hline
    \Strut$a$ & 1 & 0 & 0 & 0 & 1 & 0 & 0 & 0 \\ \hline
    \Strut$k$ & 0
    &$-1$ & 1 & 0 & 0 &$-1$& 1 & 0 \\ \hline
\end{tabular}
\end{center}
\caption{Canonical choices of the symmetry parameters $a,f$ and $k$.}
\label{tab:symmpar}
\end{table}

We will consider perfect fluid models. The energy momentum tensor,
${\tilde T}_{ab}$, is thus given by
\be
  {\tilde T}_{ab} = {\tilde \mu}u_a u_b + {\tilde p}(u_a u_b + g_{ab})\ ,
\ee
where $\tilde{\mu}$ is the energy density; $\tilde{p}$ is the pressure; and
$u^a$ the 4-velocity of the fluid.
We will assume
\be
   \tilde{p}=\left( \gamma-1 \right)\tilde{\mu}\
\ee
as an equation of state where the parameter $\gamma$ takes values in the
interval $1 \leq \gamma < 2$, which includes dust ($\gamma = 1$) and
radiation ($\gamma = 4/3$). Thus we have excluded the value $\gamma = 2$,
which corresponds to a stiff fluid.
The reason for
this is that the corresponding models behave quite
differently compared
to those in the interval $1 \leq \gamma < 2$, and thus need special
treatment. Note that all dust models are known \cite{kra}.
Note also that all self-similar LRS
models are of Petrov type D (or 0) and that
the magnetic part of the Weyl tensor is zero.

The outline of the article is the following: In section 2 we
rewrite the field equations in two steps. We first express the field
equations in a set of variables associated with the normal congruence
of the symmetry surface. We then introduce
a dimensionless set of variables in order to obtain a maximal
reduction of the coupled system of ordinary differential equations.

In section 3 the reduced phase spaces and invariant submanifolds of the
various models are discussed. The dimensionality of the fully
reduced phase spaces together with
the relation between the models, in terms of Lie contractions, are given in a
diagram. Similarities and differences between the various models are
discussed. The relation between the field equations for the SSS models and
the timelike-self-similar (TSS) models and the possibility of
extending the SSS models to the TSS sector is commented on.

In section 4 an equilibrium (critical, singular)
point analysis of the SSS type $_1$I and $_f$V
LRS models is carried out.
Asymptotic expressions for the line element and the kinematic fluid quantities
are given.
The remaining SSS and TSS LRS models
need special treatment, and will be discussed elsewhere.

In section 5 complete phase portraits are given for a number of
type $_1$I and $_f$V models.
Some global issues are also discussed.
In section 6 we discuss the nature of the singularities occurring in the
SSS type $_1$I and $_f$V models.
Appendix A describes the properties of the fluid congruence
and the condition for the spacetimes to belong to Petrov type 0.
Appendix B gives the relation between the presently used coordinates
and those of the fluid approach.

\section{Derivation of the autonomous DE}

Misner has introduced a useful metric parametrization in the context
of SH cosmology \cite{mis}. In the present case an analogous parametrization
can be introduced
\be
  D_1 = e^{\beta^0 -  2\beta^+}\ ,\qquad  D_2 = e^{\beta^0 + \beta^+}\ .
\ee
This parametrization is closely related to the kinematic properties
of the normal congruence of the $ds^2$ geometry (which is conformally
related to the SSS geometry).
The expansion, $\theta$, and the shear, which can be described by a
quantity $\sigma_+$, of this congruence, are related to $\beta^{0,+}$ by
\be
   \theta = 3\dot{\beta}^0\ ,\qquad \sigma_+ = 3\dot{\beta}^+\ ,
\ee
where the dot stands for $d/dt$.
The tetrad components of the fluid velocity
are conveniently parametrized by
$(1,v,0,0)/\sqrt{1 - v^2}$ where $v$ is just the 3-velocity
with respect to the symmetry surfaces.

Einstein's equations, ${\tilde G}_{ab} = {\tilde T}_{ab}$, and
${\tilde T}^{ab}{}_{;b} = 0$
lead to:
\begin{center}
  {\bf{Evolution equations}}
\end{center}
\be\eqalign{
  \dot{\theta} &=-\frac{1}{3}\theta^2 - \frac{2}{3}\sigma_+^2 +
  2\left( a+f \right)fB_1^2 - \frac{1}{2}\frac{\left( 3\gamma -2
  \right)+\left( 2-\gamma \right)v^2}{\Hfunc}\mu_n\ ,\cr
  \dot{\sigma}_+ &= -\theta\sigma_+ + 2\left[ \left( a+\frac{2}{3}f
  \right)\sigma_+ - \frac{1}{3}f\theta \right]vB_1 + 2\left( a+f
  \right)fB_1^2 - kB_2^2\ ,\cr
  \dot{B_1} &=\frac{1}{3}\left( -\theta+2\sigma_+ \right)B_1\ ,\cr
  \dot{B_2} &=-\frac{1}{3}\left( \theta+\sigma_+ \right)B_2\ ,\cr
  \dot{v} &= \frac{1-v^2}{3\gamma\left( 1-\left(
      \gamma -1 \right)v^2 \right)}\left[
      \gamma\left( \left( 3\gamma-4 \right)\theta+2\sigma_+ \right)v +
      3\left( \left( \gamma-1 \right)\left( 2f-\left( 2a+3f \right)\gamma
      \right)v^2 +\left( 2-\gamma \right)f\right)B_1 \right]\ ,\cr
}\ee
where
\be
   B_1 = e^{-\beta^0+2\beta^+} = D_1{}^{-1}\ ,\qquad
   B_2 = e^{-\beta^0-\beta^+} = D_2{}^{-1}\ .
\ee
The quantity
$\mu_n$ stands for the energy density of the fluid, measured by an observer
associated with the normal congruence
of the symmetry surfaces, multiplied with the factor $e^{-2fx}$. It is
related to $\tilde{\mu}$, the energy density of the fluid, by
\be
 \mu_n=\frac{1+(\gamma-1)v^2}{1-v^2}\tilde{\mu}e^{-2fx}\ .
\ee
The above evolution equations are not independent. Instead they are related
by 

\begin{center}
  {\bf{Constraint equation}}
\end{center}
\be
   \gamma v\mu_n + 2\left[ \Hfunc \right]\left[
   \left( a+\frac{2}{3}f \right)\sigma_+ -
   \frac{1}{3}f\theta \right]B_1=0\ .
\ee
\begin{center}
  {\bf{Defining equation for $\mu_n$}}
\end{center}
\be\label{eq:mun}
  \mu_n = \frac{1}{3}\left( \theta^2-\sigma_+^2 - 3 \left( 3a+f
   \right)\left( a+f \right)B_1^2 + 3kB_2^2 \right)\ .
\ee

Of interest is also:
\begin{center}
  {\bf{Auxiliary equation}}
\end{center}
\be
  \dot{\mu}_n=\frac{\gamma\mu_n}{\Hfunc}\left(
  -\theta+2(a+f)B_1v+\frac{1}{3}\left( 2\sigma_+-\theta \right)v^2
  \right)\ .
\ee

Note the close relationship between the above presentation of the
field equations and the one given by Hewitt and Wainwright for the
SH type V case \cite{hew}.
This close connection will allow us
to transfer some of the ideas used by Hewitt and Wainwright to the present
class of models. In particular, we will follow their treatment of the
constraint and the introduction of $\theta$-normalized dimensionless
variables:
\be
   \Sp = \frac{\sigma_+}{\theta}\ ,\qquad A = \frac{B_1}{\theta}\ ,\qquad
    K = \left (\frac{B_2}{\theta} \right)^2\ .
\ee

The density $\mu_n$ is replaced by the density parameter $\Omega_n$, which
is defined by
\be
  \Omega_n=\frac{3\mu_n}{\theta^2}\ .
\ee
An introduction of a dimensionless time variable $\tau$,
\be
  \frac{dt}{d\tau}=\frac{3}{\theta}\ ,
\ee
leads to a decoupling of the $\theta$-equation
\be
  \frac{d\theta}{d\tau}=-\left( 1+q \right)\theta\ ,
\ee
where the deceleration parameter $q$, associated with the normal
congruence of the SH $ds^2$ geometry, is given by
\be
  q=2\Sp^2 - 6(a+f)fA^2 +
  \frac{1}{2}\frac{ (3\gamma-2)+(2-\gamma)v^2}{1+(\gamma - 1)v^2}\Omega_n\ .
\ee
Alternatively one can write $q$ as
\be
  q=2-6\left( a+f \right)\left( 3a+2f \right)A^2 + 6kK -
  \frac{1}{2}\frac{\left( 3\left( 2-\gamma \right)+\left( 5\gamma - 6
  \right)v^2 \right)}{\Hfunc}\Omega_n\ ,
\ee
by using the definition of $\Omega_n$. The remaining
evolution equations can now be
written in dimensionless form:
\begin{center}
  {\bf{Evolution equations}}
\end{center}
\be\eqalign{
  \Sigma_+^{\prime} &= \left[ -2+q+2vA\left( 3a+2f \right)
  \right]\Sigma_+ - 2fvA + 6fA^2\left( a+f \right) - 3kK\ ,\cr
  A^{\prime} &= \left( 2\Sigma_++q \right)A\ ,\cr
  K^{\prime} &= 2\left( -\Sigma_++q \right)K\ ,\cr
  v^{\prime} &= \frac{1-v^2}{\gamma\left( 1-\left( \gamma -1 \right)v^2
  \right)}\left[ \gamma\left( \left( 3\gamma-4 \right)+2\Sigma_+ \right]v +
  3\left( \left( \gamma-1 \right)\left( 2f-\left( 2a+3f \right)\gamma
  \right)v^2 +\left( 2-\gamma \right)f\right)A \right]\ ,\cr
}\ee
where a prime denotes $d/d\tau$.
\begin{center}
  {\bf{Constraint equation}}
\end{center}
\be\label{eq:constraint}
  \gamma v \Omega_n + 2\left[ \Hfunc \right]\left[ \left( 3a+2f
  \right)\Sigma_+ - f \right]A = 0\  .
\ee
\begin{center}
  {\bf{Defining equation for $\Omega_n$}}
\end{center}
\be\label{eq:omega}
  \Omega_n = 1 - \Sigma_+^2 - 3\left( 3a+f \right)\left( a+f
  \right)A^2 + 3kK\ .
\ee
\begin{center}
  {\bf{Auxiliary equation}}
\end{center}
\be
  \Omega_n^{\prime} = \frac{\Omega_n}{1+(\gamma -1)v^2}
  \left[ -(3\gamma -2)+2q+6\gamma (a+f)vA +
  \left(2\gamma\Sigma_++2q(\gamma -1)-(2-\gamma)\right)v^2\right]\ .
\ee

The above equation system exhibits a discrete symmetry.
The field equations are invariant under the transformations
\be\label{eq:disc}
  (\Sigma_+,A,K,v) \rightarrow (\Sigma_+,-A,K,-v)\ ,\ .
\ee

The line element can be obtained when $K, A$ and $\theta$ has been found
through the relations
\be
  D_1^2 = (\theta A)^{-2}\ ,\qquad   D_2^2 = (\theta^2 K)^{-1}\ ,\qquad
  t = 3\int \frac{d\tau}{\theta}\ .
\ee
The relationship between the kinematic fluid quantities and the
$\theta, K, A, \Sigma_+, v$ variables is given in Appendix A.

\section{Reduced phase spaces and invariant submanifold structure}

The $K$-equation decouples for the type $_1$I and $_f$V models leaving a
reduced system of equations for the $\Sigma_+,A,v$ variables, related by the
constraint in eq.(\ref{eq:constraint}).
The relationship between the various models is given in terms of
Lie contractions in Diagram 1 together with the dimension of the
reduced phase space.

\begin{figure}[ht]
\begin{center}
{\tt    \setlength{\unitlength}{0.92pt}
\begin{picture}(213,270)
  \thinlines
              \put(-15,74){\line(1,0){211}}
              \put(-15,76){\line(1,0){211}}
              \put(11,144){\line(1,0){185}}
              \put(-15,183){\line(1,0){211}}
              \put(-15,185){\line(1,0){211}}
              \put(11,224){\line(1,0){185}}
              \put(-15,260){\line(1,0){211}}
              \put(33,295){\line(0,-1){260}}
              \put(-15,74){\line(0,-1){40}}
              \put(-15,35){\line(1,0){30}}
              \put(-15,260){\line(1,0){30}}
              \put(-15,260){\line(0,-1){75}}
              \put(-15,183){\line(0,-1){107}}

              \put(18,238){1}
              \put(18,202){2}
              \put(18,160){3}
              \put(18,105){2}
              \put(18,52){1}

              \put(124,88){\vector(0,-1){28}}  
              \put(57,117){\vector(3,-1){55}}  
              \put(45,131){\vector(0,1){67}}   
              \put(100,173){\vector(0,1){25}}  
              \put(155,173){\vector(0,1){25}}  
              \put(100,215){\vector(0,1){20}}  
              \put(55,212){\vector(3,2){37}}   
              \put(145,210){\vector(-3,2){40}} 
              \put(100,158){\vector(1,-3){18}} 
              \put(152,158){\vector(-1,-3){18}}

              \put(44,275){\large{{\rm{Models - Lie contractions}}}}
              \put(13,270){Dim}
              \put(-8,220){HH}
              \put(-8,52){HH}
              \put(-10,125){HSS}
              \put(120,48){$I$}
              \put(118,92){$\mbox{}_1I$}
              \put(142,162){$\mbox{}^*KS$}
              \put(149,201){$KS$}
              \put(87,162){$\mbox{}_1^*III$}
              \put(93,201){$III$}
              \put(42,201){$V$}
              \put(41,120){$V_f$}
              \put(98,239){$I$}
              \put(11,35){\framebox(185,260){}}
\end{picture}}
\caption{Lie contractions for self similar and spatially homogeneous LRS
          models. The abbreviations Caus., HH, HSS stand for causal
          character, hypersurface homogeneous and hypersurface self-similar
          respectively.}
\end{center}
\end{figure}
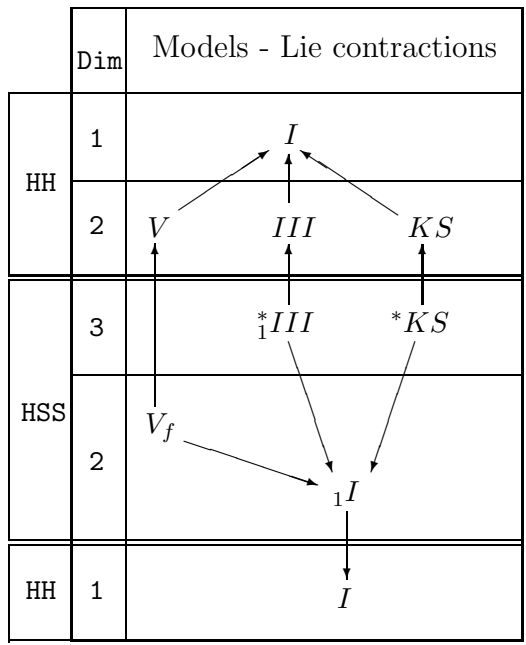

We have chosen to present the equation in expansion normalized variables
since this leads to relatively simple equations. However, a very
desirable property is compactness of the reduced phase space. In this context
eq.(\ref{eq:mun}) plays an essential role. One can use the fact that
${\mu}_n$ is nonnegative to produce inequalities which can be used in
order to find ``dominant'' quantities. The expansion, $\theta$, is
such a quantity for {\it some\/} of the models (those for which
$k \leq 0$, $(3a + f)(a + f) \geq 0$; as seen by eq.(\ref{eq:mun})).
Note that $\theta$ cannot change sign for the matter filled
``$\theta$-dominated'' models because of eq.(\ref{eq:mun}).
Furthermore, because of the discrete symmetry in eq.(\ref{eq:disc})
one can therefore, without loss of generality,
assume $A \geq 0$ (this just corresponds to looking at models with positive
expansion). Unfortunately one cannot make this assumption for the other
models where $\theta$, and thus also $A$, may change sign.

The boundary consists of a number of invariant subsets which plays an
important role when determining the qualitative properties of the orbits
in the reduced phase space. The boundary of the reduced phase space
will thus be included. We have the
following invariant submanifolds on the boundary:
(i) $A=0$, (ii) $v=1$, (iii) $v=-1$, (iv) the vacuum submanifold
$1 - \Sigma_+^2 - 3(3a+f)(a+f)A^2 = 0$. For the $k \neq 0$ models we also
have (v) $K=0$ as an invariant submanifold of the reduced phase space.

The $v= \pm 1$ submanifolds
has a physical interpretation in terms of models with directed fluxes
of neutrinos (see \cite{bog}).
In the vacuum submanifold case one can interpret
$v$ as the velocity of a test fluid (see \cite{ship}).
The constraint leads to $v\Omega_n = 0$ when $A=0$. For $v=0$ we obtain
the reduced equations for the orthogonal SH LRS type I, III and KS models.
When $\Omega_n =0$ one obtains the
reduced equations for the vacuum models for the same models
but with a test fluid.
For the $k\neq 0$ models, the $K=0$ manifold yields the same equations
as the reduced type $_1$I equations.
Note that $v = 0$, in contrast to the SH models, is not an invariant
submanifold (except for $\gamma = 2$). Thus one naturally obtains,
for the present class of LRS models, the well known result that
SSS models are tilted if $\gamma \neq 2$
(i.e., there are no SSS models with a fluid flow
orthogonal with respect to the symmetry surfaces) \cite{wu}.

\subsection{The reduced phase space of type $_f$V and $_1$I models}

For these models $k=0$. This results in a decoupling of the equation for $K$.
Since $K$ decouples the boundary value $K=0$ is of no interest (it is enough
to analyze the reduced system of equations). Thus
$K$ satisfy the inequality $K>0$. This leaves a
coupled system of equations for the remaining variables $(\Sp,A,v)$,
related by the constraint equation.
As mentioned above, the $(3a + f)(a + f) = 0$ and the
$(3a + f)(a + f) < 0$ models, do not have a compact reduced phase
space. Moreover, the $3a + f = 0$ and $a + f = 0$ models behaves drastically
different compared to the other models. This is seen in the equilibrium point
analysis below. Shikin has been able to find the general
solution for the $3a + f$ and $a + f = 0$ cases \cite{ship,shigrg}.
Shikin has also qualitatively investigated the
$(3a + f)(a + f) < 0$ models \cite{ship}.
Below we will focus on the ``$\theta$-dominant''
$(3a + f)(a + f) > 0$ case which contains the SH type V models
and the SSS type $_1$I models.
Recall that one can assume that $A \geq 0$ for the $\theta$-dominant
models. However, we will also list the equilibrium points for the remaining
models since this gives an indication of how different the three types of
models ($(3a + f)(a + f) < 0, (3a + f)(a + f) =0, (3a + f)(a + f) > 0$) are.

There are four invariant submanifolds describing the boundary of the
reduced equation system of the
$\theta$-dominant models: The $v=\pm1$ submanifolds;
the vacuum submanifold $1 - \Sigma_+^2 - 3(3a+f)(a+f)A^2 = 0$,
and the $A=0$ submanifold which can be divided into three parts depending
on $v < 0, v = 0, v > 0$.

\subsection{The reduced phase space of type $^*_1$III and $^*$KS models}

It follows from
eq.(\ref{eq:mun}) that one obtains a compact reduced phase space
in the (SSS) type $^*_1$III models. However, the $\theta$
normalized variables are not compact for the $^*$KS models.
In a forthcoming article we will
show how one can use eq.(\ref{eq:mun}) to produce compact reduced
phase space variables in the $^*$KS case.
The type $^*_1$III models are characterized by the choices
$k = -1,a = 0,f = -1$.
The reduced system consists of equations for the $\Sigma_+,A,K,v$ variables,
related by the constraint in eq. (\ref{eq:constraint}).
Here the $K=0$ value is included. This leads to the
inequality $K \geq 0$. The boundary is described by five pieces:
The $v=\pm1$ submanifolds;
the vacuum submanifold $1 - \Sigma_+^2 - 3A^2 - 3K= 0$;
the $A=0$ submanifold, and the $K=0$ submanifold.

\subsection{The relation between the SSS and TSS field equations}

It is important to note that the above equation system (where $\mu_n$
($\Omega_n$) is
assumed to be eliminated by use of eq.(\ref{eq:mun}) (eq.(\ref{eq:omega})))
also describes the field equations
in the TSS region if one makes the change $K \rightarrow -K$
in the $k\neq 0$ case. However, in
the TSS case $v^2 \geq 1$ instead of, as presently, $v^2 \leq 1$.

It's natural to extend
solutions to the TSS sector when the symmetry surface changes causality and
when one does not run into a (non-coordinate) singularity
(compare with the LRS type V discussion by Collins and Ellis \cite{coel}).
A number of phenomena occurs in the TSS region which do not
exist in the SSS case. Hence, we will investigate the TSS region separately
in a forthcoming article. This will allow us
to obtain a more global picture of the self-similar LRS models
than in the present article which will, from now on, deal exclusively
with the SSS case.

\section{Equilibrium points and asymptotic analysis}

The reduced phase space is determined by a coupled system
$d{\bf X}/d{\tau} = {\bf F}({\bf X})$ subject to a constraint $G({\bf X})=0$
given by eq.(\ref{eq:constraint}) ({\bf X} constitute the reduced phase
space variables).
Of central importance to the investigation of the dynamical system are
the equilibrium points which are determined by the equations
${\bf F}({\bf X})=0, G({\bf X})=0$. Below
the equilibrium points are given together
with $\Omega_n$, which shows if
an equilibrium point is located on a vacuum submanifold or not.
The gradient of $G({\bf X})$, which is used to locally solve the constraint
to linear order, and the eigenvalues and eigenvectors of the remaining
locally unconstrained system are then listed.

A local analysis of the equilibrium points is then used to express
the {\it fluid\/} energy density ($\tilde \mu$), expansion ($\tilde \theta$),
shear scalar ($\tilde \sigma$), acceleration ($\tilde a$),
divergence of the acceleration (${\tilde a}^\mu{}_{;\mu}$);
the line element ($D_{1,2}$); and
the Weyl scalar ($C$), for the various asymptotes in the
interior matter filled phase space.
The tables describing asymptotic properties
differ from the ones given by Collins and Ellis for the SH type V models
\cite{coel}. Their tables
relate the initial and final state of a given orbit (or orbits).
Such tables will be given in the next section when we discuss global
behaviour. Our tables also differ from those in reference
\cite{coel} in that we express the behaviour in terms of the
independent variable $\tau$ instead of the proper time associated with the
comoving fluid frame. The reason for this is that the SSS and SH models
need separate treatment if one wants to use proper fluid time.
The results in this section and those by \cite{coel,col} for SH type
V models are easily related by using, e.g., $\tilde \theta$, as independent
variable. We only give the $\tau$-dependence; the self-similar
dependence on $e^{fx}$ is given in Appendix A.
The acceleration scalar, $\tilde a$, and the divergence of the acceleration,
${\tilde a}^\mu_{;\mu}$, are of course zero for the dust, $\gamma = 1$,
models.

\subsection{Type $_f$V and $_1$I models}

The reduced phase space consists of the constraint surface in the
$(\Sigma_+,A,v)$ space, subject to the inequalities
$\Omega_n \geq 0, v^2 \leq 1$. For the $\theta$-dominated models we also have
$A\geq 0$.

\subsubsection{Equilibrium points with zero tilt, $v=0$}

\eqp{$F$}                 
\be\eqalign{
  \Sp &= 0\ ,\quad A = 0\ ;\quad \Omega_n = 1\ ;\cr
  \nabla G &= \left( 0,-2f,\gamma\right)\ ,\quad
  (v \, \elim)\ ;\cr
  {} {} & -\fr32 (2 - \gamma)\ ,\quad (1,0)\ ;\quad
          \fr12 (3\gamma - 2)\ ,\quad (0,1)\ ,\cr
}\ee
for $1\leq\gamma<2$,
where the constraint has been used to solve for the $v$ variable and
where the eigenvalues of the resulting unconstrained system
are grouped together with the corresponding eigenvectors.
The equilibrium point is a saddle point. In Table 2
we give asymptotic
expressions for the single initial asymptote entering the interior of the
reduced physical phase space of the type $_f$V and $_1$I models.

\def\Strut{\vrule width 0pt height 13pt depth 6pt{}}
\begin{table}[h]
\begin{center}
  \begin{tabular}{|c|c|c|c|c|c|c|c|c|c|}  \hline 
     \Strut As.\,St. & $\tilde{\mu}$ & $\tilde{\theta}$ & $\tilde{\sigma}$
   & $\tilde{a}$ & $\tilde{a}_{;\mu}^{\mu}$ & $D_1$ & $D_2$ & $ C $ \\ \hline
   \Strut${\rm i}$ & $e^{-3\gamma\tau}$ & $e^{-3\gamma\tau /2}$ &
   $e^{(3\gamma-4)\tau /2}$ & $fe^{(\gamma-1)\tau}$ &
   $fe^{(3\gamma-4)\tau}$ &
   $ e^{\tau}$ & $e^{\tau}$ & $fe^{-2\tau}$ \\ \hline
\end{tabular}
\end{center}
\caption{Asymptotic expressions for the single initial asymptote coming
         from the equilibrium point $F$. Here, and below As. St. is an
        abbreviation for asymptotic state. Initial states are denoted by i
         and final states by f.}
\label{tab:F}
\end{table}

\noindent
{\it The equilibrium points\/} $K^0_{\pm}$:    

\be\eqalign{
  \Sp &=\pm 1\ ,\qquad A=0\ ;\qquad \Omega_n = 0\ ;\cr
  \nabla G &= \left( 0,2\left(3a+f  \right),0 \right)\ ,
  \quad {\rm for} \quad \Sp=+1\ ,\cr
  \nabla G &= \left( 0,-6\left(a+f  \right),0 \right)\ ,
  \quad {\rm for} \quad \Sp=-1\ ;\quad (A \, \elim)\ ;\cr
  & {} {} 3(2 - \gamma)\ ,\quad (1,0)\ ;\quad 2\Sp+(3\gamma - 4)\ ,
       \quad (0,1)\ ,\cr
}\ee
for $1 \leq \gamma < 2$.
Different equilibrium points are often closely related through sign changes
or zero values of $v$ and $\Sigma_+$. In such cases we denote the equilibrium
points with a common kernel and indicate the value of $v$ in the
upper position and of $\Sigma_+$ in the lower (if there only exists points
with a given value of $v$ or $\Sigma_+$ we do not indicate the corresponding
value). Thus in the present case
of $K^0_{\pm}$, 0 refers to $v=0$ and $\pm$ refers to the sign of $\Sigma_+$.
(Note that this leads to a somewhat different notation than that used
by Hewitt and Wainwright for the SH type V case \cite{hew}.)
The point $K^0_-$ is a saddle point on the boundary and does not give
rise to any asymptotes in the interior phase space. The point $K^0_+$
on the other hand is a stable source giving rise to a 1-parameter set
of asymptotes described in Table 3.

\def\Strut{\vrule width 0pt height 13pt depth 6pt{}}
\begin{table}[h]
\begin{center}
  \begin{tabular}{|c|c|c|c|c|c|c|c|c|c|}  \hline 
  \Strut As.\,St. & $\tilde{\mu}$ & $\tilde{\theta}$ & $\tilde{\sigma}$
   & $\tilde{a}$ & $\tilde{a}_{;\mu}^{\mu}$ & $D_1$ & $D_2$ & $ C $ \\ \hline
 \Strut${\rm i},\Sp=+1$ & $e^{-3\gamma\tau}$ &
 $e^{-3\tau}$ & $e^{-3\tau}$ & $e^{(3\gamma-5)\tau}$ &
 $e^{2(3\gamma-5)\tau}$ & $e^{-\tau}$& $e^{2\tau}$ &
 $e^{-6\tau}$\\ \hline
\end{tabular}
\end{center}
\caption{Asymptotic expressions for the fluid congruence around
the $K^{0}_+$-point.}
\label{tab:K0p}
\end{table}

\eqp{$M^0$}                

This point only exists for $f=0$ and therefore we can choose $A>0$.
\be\eqalign{
  \Sp &= 0\ ,\quad A = \frac{1}{3a}\ ;\quad \Omega_n = 0\ ;\cr
   \nabla G &= (2,0,0)\ ;\quad (\Sp \, \elim)\ ;\cr
   {} {} & -(3\gamma - 2)\ ,\quad (1,0)\ ;\quad 3\gamma - 4\ ,\quad (0,1) .
}\ee
There exists a bifurcation for $\gamma=4/3$, which needs special
consideration; we refer to \cite{hew}. For $1 \leq \gamma < 4/3$ the point
is stable leading to a 1-parameter set of final asymptotes, among them
the Friedmann asymptote. For $4/3 \leq \gamma < 2$ the point is a saddle
and has no asymptotes into the physical phase space, except for the special
Friedmann asymptote. The asymptotic behaviour is described in Table 4.

\def\Strut{\vrule width 0pt height 13pt depth 6pt{}}
\begin{table}[h]
\begin{center}
  \begin{tabular}{|c|c|c|c|c|c|c|c|c|c|}  \hline 
 \Strut As.\,St. & $\tilde{\mu}$ & $\tilde{\theta}$ & $\tilde{\sigma}$
 & $\tilde{a}$ & $\tilde{a}_{;\mu}^{\mu}$ & $D_1$ & $D_2$ & $ C $ \\
 \hline     \Strut${\rm f}, \gamma>4/3$ & $e^{-3\gamma\tau}$ & $e^{-\tau}$ &
  $0$ & $0$ & $0$ & $e^{\tau}$& $e^{\tau}$ & $0$\\ \hline
 \Strut${\rm f}, \gamma<4/3$ & $e^{-3\gamma\tau}$ & $e^{-\tau}$ &
 $e^{(3\gamma-5)\tau}$ & $e^{(3\gamma-5)\tau}$ &
 $e^{-3(2-\gamma)\tau}$ & $e^{\tau}$& $e^{\tau}$ & $e^{-4\tau}$\\ \hline
\end{tabular}
\end{center}
\caption{Asymptotic expressions for the fluid congruence around the
$M^0$-point.}
\label{tab:M0}
\end{table}

\subsubsection{Points with intermediate tilt, $0<v^2<1$}

{\it The equilibrium points\/} $M^{v_\pm}$:
$1< \gamma < 2$; $\gamma \neq 2f/(2a + 3f)$.  
\be\eqalign{
  \Sp &= \frac{f}{3a+2f}\ ,\quad A=\frac{\epsilon}{3a+2f}\ ,\quad
   v = \frac{\epsilon \gamma c_2}{2c_1(\gamma-1)}\pm \sqrt{\frac{\gamma^2
       c_2^2 + 4(2-\gamma)(\gamma-1)fc_1}{4c_1^2(\gamma-1)^2}}\ ;\quad
   \Omega_n = 0\ ;\cr
  c_1 &:= (3\gamma-2)f+2a\gamma\ ,\quad
  c_2 := 2(\gamma-1)f + (3\gamma-4)a\ ;\cr
  \nabla G &= \left( \frac{2\epsilon v^2 (3a+2f)(\gamma-1) - 2f\gamma v +
             2\epsilon (3a+2f)}{3a+2f},
             -\frac{6\epsilon \gamma v(3a+f)(a+f)}{3a+2f},0 \right)\ ;\cr
      {} & (A \, {\elim})\ ;\quad
      -\frac{3(1-\epsilon v)\left(c_1v-\epsilon
    (2-\gamma)f \right)}{(3a+2f)\gamma v}\ ;\qquad \frac{\epsilon e_1 v + e_2}
        {(\gamma-1)^2c_1^3}\ ,\quad (0,1)\cr {}
  e_1 &:= -3\epsilon((2(\gamma-1)
        (2-\gamma)^2f^2-c_2^2f\gamma^3)(\gamma-1)c_1^2
          +2(4(\gamma-1)^2-(2-\gamma))(\gamma-1)^2c_1^3f \cr
          & +4(\gamma-1)^3a\gamma c_1^3
          -4(\gamma-1)(\gamma-2)c_1c_2^2f\gamma^2+c_2^4\gamma^4)\ , \cr
e_2 &:= -3( (3(\gamma-2)f+c_1\gamma)(\gamma-1)(\gamma-2)c_1 f
          -(\gamma-1)^2c_1^3-(\gamma-2)c_2^2f\gamma^2)c_2\gamma\ .
}\ee
The $\pm$ in $M^{v_\pm}$ refers to the different signs of the root
in the expression for $v$. The condition $\gamma \neq 2f/(2a + 3f)$
corresponds to $c_1 \neq 0$.
The eigenvector of the second eigenvalue above points in the direction of the
vacuum submanifold, while the eigenvector of the first eigenvalue points into
the physical phase space. The expression for the eigenvector associated
with the first eigenvalue is quite cumbersome and will not be given.
Note that there are two sets of points for the
non-$\theta$-dominated models since $A$ can be both positive and negative
in that case. For the $\theta$-dominated models there is only a single
pair of points. The requirement of $A\geq0$ taken
together with the sign of $3a + f$ determines
the sign of the parameter $\epsilon$ which can take the values $\pm 1$.
The first eigenvalue is negative for the $\theta$-dominated models.
Hence it follows that, in the $\theta$-dominated
cases, possible initial asymptotes
lies in the vacuum submanifold. Thus
the only interior asymptote is a final one. Its generic behaviour is
given in Table 5.

\def\Strut{\vrule width 0pt height 13pt depth 6pt{}}
\begin{table}[h]
\begin{center}
  \begin{tabular}{|c|c|c|c|c|c|c|c|c|}  \hline 
 \Strut As.\,St. & $\tilde{\mu}$ & $\tilde{\theta}$ & $\tilde{\sigma}$
   & $\tilde{a}$ & $\tilde{a}_{;\mu}^{\mu}$ & $D_1$ & $D_2$ & $ C $ \\ \hline
    \Strut${\rm f}$ & $e^{-\lambda^2\tau}$ & $e^{-\Gamma_1\tau}$ &
$e^{-\Gamma_1\tau}$ &
        $e^{-\Gamma_1\tau}$ &$e^{-2\Gamma_1\tau}$ & $e^{\Gamma_1\tau}$&
  $e^{\Gamma_2\tau}$ & $e^{-\lambda^2}\tau$\\ \hline
\end{tabular}
\end{center}
\caption{Asymptotic expressions for the fluid congruence around the
$M^{v_{\pm}}$-points. The constants $\Gamma_{1,2}$ are given by
$\Gamma_1=\frac{3a}{3a+2f}$ and $\Gamma_2=\frac{3(a+f)}{3a+2f}$.
Here and below $\lambda^2$ will stand for a constant which will not be given
because it is quite complicated (the actual value of $\lambda^2$ is
different for different tables).}
\label{tab:Mv}
\end{table}

\noindent
{\it The equilibrium points\/} $M^{v_\pm}$: $\gamma = 1$.

\be\eqalign{
  \Sp &= \frac{f}{3a+2f}\ ,\quad
  A = \pm \frac{1}{3a+2f}\ ,\quad v=\pm \frac{f}{a}\ ; \quad \Omega_n =0\ ;\cr
   \nabla G &=\left( \pm \frac{2\left( 3a-f \right)\left( a+f \right)}{a\left(
   3a+2f \right)}, -\frac{6f\left(3a+f \right)\left( a+f
   \right)}{a\left( 3a+2f \right)},0\right)\ ;\cr
   {} & (A \, {\elim})\ ,
}\ee
The local analysis yields a single multiple valued
eigenvalue ($m$ below) and a linear solution of the following form:
\be\eqalign{
  \Sp &= \frac{f}{3a+2f} + k_2e^{m\tau}\ , \qquad v=\pm\frac{f}{a} + \left(
  k_1+k_2p\,\tau\right)e^{m\tau}\ ; \cr
  m &:=-\frac{3(a+f)(a-f)}{(3a+2f)a}\ ,
  p := \frac{3f(3a+2f)(a-f)(a+f)^2}{(3a-f)a^3}\ ,
}\ee
where $k_1$ and $k_2$ are constants. For $\theta$-dominated models with
$(3a+f)<0$ and models with $(a+f)>0$ and $a-f<0$ none of these points exist
in the physical phase space. However, for models with $(a+f)>0$ and $a-f>0$ the
point $M^{v_+}$ exists and is a sink. For non-$\theta$-dominated models
it can be a sink or a source depending on the values of $a$ and $f$.
Asymptotic behaviour
is given in Table 6.

\def\Strut{\vrule width 0pt height 13pt depth 6pt{}}
\begin{table}[h]
\begin{center}
  \begin{tabular}{|c|c|c|c|c|c|c|c|c|}  \hline 
 \Strut As.\,St. & $\tilde{\mu}$ & $\tilde{\theta}$ & $\tilde{\sigma}$
   & $\tilde{a}$ & $\tilde{a}_{;\mu}^{\mu}$ & $D_1$ & $D_2$ & $ C $ \\ \hline
    \Strut${\rm f}$ & $e^{-\lambda_1^2\tau}$ & $e^{-\Gamma_1\tau}$ &
$e^{-\Gamma_1\tau}$ &
        $0$ &$0$ & $e^{\Gamma_1\tau}$&
  $e^{\Gamma_2\tau}$ & $e^{-\lambda^2\tau}$\\ \hline
\end{tabular}
\end{center}
\caption{Asymptotic expressions for the fluid congruence around the
$M^{v_{\pm}}$-points for $\gamma=1$. The constants $\Gamma_{1,2}$ are given by
$\Gamma_1=\frac{3a}{3a+2f}$ and $\Gamma_2=\frac{3(a+f)}{3a+2f}$ while
$\lambda_1^2 = \frac{3(2a^2 - f^2)}{a(3a + 2f)}$.}
\label{tab:Mvg1}
\end{table}

\noindent
{\it The equilibrium points\/} $M^{v_\pm}$: $\gamma = 2f/(2a+3f)$.
\be\eqalign{
 \Sp &= \frac{f}{3a+2f}\ ,\quad A = \pm\frac{1}{3a+2f}\ ,\quad
   v = \pm\frac{2a+3f}{4a+f}\ ; \quad \Omega_n = 0\ ;\cr
  \nabla G &= \left(
  \pm\frac{4\left( 6a^2+2af-3f^2 \right)(3a+f)}{(4a+f)^2(3a+2f)},
  -\frac{12f(3a+f)(a+f)}{(4a+f)(3a+2f)},0 \right)\ ;\cr
   {} & (A \, {\elim})\ ;\quad
   \frac{12(a+f)(a-f)}{(3a+2f)(2a+3f)}\ ;\quad
   -\frac{6(4a+f)(a+f)(a-f)}{\left( 5a^2+4af+f^2 \right)(2a+3f)}\ .\cr
}\ee
This point only exists for models which are non-$\theta$-dominated, and
therefore we have only given the eigenvalues and not the eigenvectors.
We have also refrained from giving a table with the asymptotic
properties.

\subsubsection{Points with extreme tilt, $v^2=1$}

{\it The equilibrium points\/} $M^{\pm}$:
\be\eqalign{
  \Sp &= \frac{f}{3a+2f}\ ,\quad A = \mp \frac{1}{3a+2f}\ ,
  \quad v = \pm 1\ ;\quad \Omega_n = 0\ ;\cr
  \nabla G &= \left( \pm\frac{6\gamma(a+f)}{3a+2f},
              \frac{6\gamma(a+f)(3a+f)}{3a+2f},0 \right)\ ;\quad
  (\Sp \, {\elim})\ ;\cr
  {} & -\frac{12(a+f)}{(3a+2f)}\ ,\quad (1,0)\ ;\quad
       -\frac{6(a+f)(5\gamma-6)}{(3a+2f)(2-\gamma)}\ ,\quad (0,1)\ .\cr
}\ee
For $\theta$-dominated models with $(3a+f)<0$ only $M^+$ exists, while for
models with $(a+f)>0$ only $M^-$ exists.
The quotient $(a+f)/(3a+2f)$, appearing in both eigenvalues, is positive for
all $\theta$-dominated models. However, for non-$\theta$-dominated
models different signs are possible and thus this factor will affect
the stability of the point in this case.
For $\theta$-dominated models,
the only  bifurcation is for $\gamma=6/5$. For $\gamma < 6/5$ the existing
point is a saddle and there is no asymptote from or into the interior region
of the phase space, while
for $\gamma>6/5$, the points are always sinks. Note that the fluid
quantities require separate treatment when $\gamma = 4/3$, even though
there is no bifurcation (note the asymptotic behaviour of
$\tilde{\sigma}$ in Table 7).

For $\gamma > 6/5$, the sink associated with the existing point does not
correspond to a curvature singularity. For $\theta$-dominated
models with $(3a+f)<0$ there is no final singular behaviour at all.
This is also true for the case $(a+f)>0$ and
$f\leq-2a(3\gamma-4)/(5\gamma-6)$; but if $f<-2a(3\gamma-4)/(5\gamma-6)$, then
the point corresponds to a crushing singularity (i.e., the kinematic
quantities blow up while the curvature stays finite (it goes to zero)).

In the above discussion about singularities,
we should really only say that the curvature and/or
kinematic quantities blow up asymptotically since we have not
investigated if one really reaches the asymptotic stage
in a finite amount of proper fluid time or
in a finite affine distance along a geodesic.
Such an investigation will not be undertaken in
the initial and final asymptotic analysis below either.
To prove that we really have singularities we would need to do an
additional analysis involving asymptotic coordinate variable transformations
for each case. However,
for the closely related SH
case and for the unphysical SH geometry associated with the SSS geometry,
the initial and final
states corresponding to possible singularities are reached
in a finite time. Hence we believe that these states correspond to
singularities and we will thus refer to them as such.

\def\Strut{\vrule width 0pt height 13pt depth 6pt{}}
\begin{table}[h]
\begin{center}
  \begin{tabular}{|c|c|c|c|c|c|c|c|c|}  \hline 
\Strut As.\,St. & $\tilde{\mu}$ & $\tilde{\theta}$ & $\tilde{\sigma}$
   & $\tilde{a}$ & $\tilde{a}_{;\mu}^{\mu}$ & $D_1$ & $D_2$ & $ C $ \\ \hline

   \Strut${\rm f}, \gamma>6/5, \gamma \neq 4/3$ & $e^{\Gamma_2\tau}$ &
   $e^{\Gamma_1\tau}$ & $e^{\Gamma_1\tau}$ & $e^{\Gamma_1\tau}$ &
   $e^{2\Gamma_1\tau}$ & $e^{3a\tau/(3a+2f)}$& $e^{3(a+f)\tau/(3a+2f)}$ &
   $e^{-6\tau}$ \\ \hline

   \Strut${\rm f}, \gamma>6/5, \gamma= 4/3$ & $e^{\Gamma_2\tau}$ &
   $e^{\Gamma_1\tau}$ & $e^{\Gamma_3\tau}$ & $e^{\Gamma_1\tau}$ &
   $e^{2\Gamma_1\tau}$ & $e^{3a\tau/(3a+2f)}$& $e^{3(a+f)\tau/(3a+2f)}$ &
   $e^{-6\tau}$ \\ \hline
\end{tabular}
\end{center}
\caption{Asymptotic expressions for the fluid congruence around the
$M^{\pm}$-point. The constants $\Gamma_{1,2,3}$ are given by
$\Gamma_1=\frac{3\left((a+f)(5\gamma-6)-(2-\gamma)a\right)}
{(3a+2f)(2-\gamma)}$, $\Gamma_2=-\frac{6\left((3\gamma-2)f + 2a\gamma\right)}
{(3a+2f)(2-\gamma)}$ and $\Gamma_3=-\frac{2a+f}{3a+2f}$.}
\label{tab:Mpm}
\end{table}

\noindent
{\it The equilibrium points\/} ${\cal{H}}^\pm$:
\be\eqalign{
  \Sp &= -1\pm 3\left( a+f \right)A\ ,\quad v=\pm 1\ ;\quad
  \Omega_n = \frac{2(1+\Sigma_+)(f-(3a+2f)\Sigma_+)}{3(a+f)}\ ;\cr
  \nabla G &= \left( \pm\frac{2\gamma (3a+2f-f\Sigma_+)}{3(a+f)} ,
        \pm 2\gamma\left( 3a+2f - f\Sigma_+ \right),
        \frac{2(\gamma-2)(1+\Sigma_+)\left( (3a+2f)\Sigma_+ -f \right)}{3(a+f)}
        \right)\ ;\cr
   {} & (A \, \elim)\ ;\quad 0\ ,\quad (1,0)\ ;\cr
  {} &\frac{-2(2(2a+3f)\Sp - a)}{a+f}\ ,\quad
       (-(1+\Sigma_+) \left( (3a+2f)\Sigma_+ - f\right)e_1,
       3\left( (2a+3f)\Sigma_+ -a \right)e_2)\ ,\cr
  e_1 &:= \left(2f(3\gamma-5)-3a(2-\gamma)\right)\Sigma_+^2 -
  \left(3a(3\gamma-4)+2f(3\gamma-5)\right)\Sigma_+ + 3\gamma a -
  (3\gamma-2)f\ ,\cr
   e_2 &:= 3a+2f-f\Sigma_+\ .\cr
}\ee
For $\theta$-dominated models only one of the lines ${\cal{H}^\pm}$ exists
in the physical part of the phase space.
For models with $(3a+f)<0$, only ${\cal{H}^-}$ exists,
while for models with $(a+f)>0$ only ${\cal{H}^+}$ is present. The
condition that $\Omega_n \geq 0$, limits the possible values of $\Sp$ to lie
in the interval $-1\leq \Sp \leq f/(3a+2f)$. We also note that for the models
with $(3a+f)<0$ the existing line of equilibrium points splits into two parts.
The part where $\Sp$ attain the values $-1\leq\Sp < 2a/(2a+3f)$ constitutes
a source while the part where $2a/(2a+3f)<\Sp\leq f/(3a+2f)$ is a sink.
For models with $(a+f)>0$ and $(f-a)<0$, the
entire line ${\cal{H}}^+$ is a source. However,
for models with $f\geq a$ the line
splits into two parts where the stability of the two parts is
governed by the same inequalities as for the $(3a+f)<0$ case.
The asymptotic behaviour is given in Table 8.

All $\theta$-dominated models with an initial point
on one of the lines ${\cal{H}^\pm}$ have an initial crushing singularity.
This crushing singularity is a curvature singularity if $\Gamma_1$ in
Table 8 is negative. Thus for models with $f>0$
there is an initial curvature singularity.

The part of ${\cal{H}^\pm}$ corresponding to a sink is never associated
with a final curvature singularity when it comes to $\theta$-dominated models.
If $(3a+f)<0$ and $2a/(2a+3f)\leq\Sp \leq (2a+f)/(4a+5f)$ there are no final
singularities at all, but for $(3a+f)<0$ and
$(2a+f)/(4a+5f) < \Sp\leq f/(3a+2f)$ there is a crushing singularity.
For $\theta$-dominated
models with $(a+f)>0$ and $a<f\leq 2a$, there is no final singular
behaviour at all. For $2a<f$, ${\cal{H}}^+$ again splits into two parts
whose behaviour is governed by the same inequalities as
in the ``splitted'' ${\cal{H}}^-$ case.
For non-$\theta$-dominated
models the situation, which will not be discussed further,
is more complicated.

\def\Strut{\vrule width 0pt height 13pt depth 6pt{}}
\begin{table}[h]
\begin{center}
  \begin{tabular}{|c|c|c|c|c|c|c|c|c|}  \hline 
\Strut As.\,St. & $\tilde{\mu}$ & $\tilde{\theta}$ & $\tilde{\sigma}$
   & $\tilde{a}$ & $\tilde{a}_{;\mu}^{\mu}$ & $D_1$ & $D_2$ & $ C $ \\ \hline

   \Strut${\rm i,f}$ & $e^{\Gamma_1\tau}$ & $e^{\Gamma_2\tau}$ &
   $e^{\Gamma_2\tau}$
   & $e^{\Gamma_2\tau}$ & $e^{2\Gamma_2\tau}$ & $e^{(2\Sigma_+-1)\tau}$&
   $e^{(1+\Sigma_+)\tau}$ & $e^{\Gamma_1\tau}$ \\ \hline

\end{tabular}
\end{center}
\caption{Asymptotic expressions for the fluid congruence around the
${\cal{H}}^{\pm}$-points. The constants $\Gamma_{1,2}$ are given by
$\Gamma_1=-\frac{2f(1+\Sigma_+)}{(a+f)}$,
$\Gamma_2=\frac{(5\Sigma_+-1)f-2a(1-2\Sigma_+)}{(a+f)}$.}
\label{tab:Hpm}
\end{table}

\noindent
{\it The equilibrium points\/} $K^{\pm}_{\pm}$:
\be\eqalign{
  \Sp &= \pm 1\ ,\quad A=0\ ,\quad v^2=1\ ;\quad \Omega_n = 0\ ;\cr
  \nabla G &= \left( -2\gamma {\rm sgn}v), 2\gamma\left( 3a+f \right),0
\right)\ ,
  \quad {\rm for\/}\ \Sp=1\ ,\cr
  \nabla G &= \left( -2\gamma {\rm sgn}(v), -6\gamma\left( a+f
\right),0\right)\ ,
  \quad {\rm for\/}\ \Sp=-1\ ;\cr
  {} & (A \, {\elim})\ ;\quad
  2(1 + \Sp)\ ,\quad (1,0)\ ;\quad
  -\frac{2}{2-\gamma}\left(
         2\Sp + (3\gamma-4)\right)\ ,\quad (0,1)\ .\cr
}\ee
The points $K^{\pm}_-$ constitute one of the boundaries of the physical
part of the lines ${\cal{H}^{\pm}}$.
They describe the boundary asymptotes which play an important role in
the dynamical system analysis. This
motivates their special treatment. The points $K^{\pm}_+$
have no asymptotes entering the interior region of the
phase space.

\eqp{${{\bar{\cal{M}}}^{\pm}}}$
\be\eqalign{
  \Sp &= \frac{f}{3a+2f}\ ,\quad A = \pm \frac{1}{3a+2f}\ ,
  \quad v = \pm 1\ ;\quad \Omega_n = 0\ ;\cr
  \nabla G &= \left(\mp \frac{2\gamma(3a+f)}{3a+2f},
                       -\frac{6\gamma(3a+f)(a+f)}{3a+2f},0\right)\ ;\quad
  (\Sp \, \elim)\ ;\cr
  {} {} & 0\ ,\quad (1,0)\ ;\quad \frac{6(a-f)}{3a+2f}\ ,\quad (0,1)\ .\cr
}\ee
These points are at the other end (compared to $K^{\pm}_-$) of
the lines ${\cal{H}^{\pm}}$.
The points have no asymptotes entering the interior region of the phase space.

\section{Global SSS behaviour}

\subsection{Type $_1$I models}

The type $_1$I models are defined by $a=0$ and $f=-1$ (see Table 1).
The model is $\theta$-dominated which implies that our variables leads to a
compact reduced phase space. Only the line ${\cal{H}}^-$ of the two lines
${\cal{H}}^\pm$ lies in the physical phase space. The stability of
${\cal{H}}^-$ is
governed by the sign of $\Sp$; for $\Sp<0$ it is a source, and for $\Sp>0$ it
is a sink.  Note that the requirement $\Omega_n\geq 0$ limits the range of
$\Sp$ to $-1\leq\Sp\leq 1/2$. The only bifurcation appears for $\gamma=6/5$,
when the equilibrium
point $M^{v_+}$ passes through $M^+$ and enters the physical phase space,
stabilizing $M^+$. As $\gamma$ approaches $2$, the point $M_{v_+}$ moves
along the line $M^+-M^{v_+}-\bar{M}^-$,
while $v$ approaches $0$. Thus there
are only two cases: $1 \leq \gamma \leq 6/5$ and $6/5 < \gamma < 2$.
In both cases there are no interior equilibrium points, and hence no limit
cycles. In addition, consideration of the orbits on the boundary leads to the
conclusion that there are no heteroclinic cycles (compare with
the SH type V case discussed in \cite{hew}).

\subsubsection{Models with $1 \leq \gamma \leq 6/5$}
The different orbits can be written symbolically as ${\cal{H}}^- \rightarrow
{\cal{H}}^-$, $F \rightarrow {\cal{H}}^-$ and $K_+^0 \rightarrow {\cal{H}}^-$.
The intermediate evolution of orbits which
define one-parameter families of solutions can be approximated by the
following heteroclinic sequences:
\renewcommand{\labelenumi}{(\theenumi)}
\begin{enumerate}
 \item $K_+^0 \rightarrow K_+^+ \rightarrow M^+ \rightarrow {\cal{H}}^-$
 \item $K_+^0 \rightarrow F \rightarrow {\cal{H}}^-$
 \item ${\cal{H}}^- \rightarrow K_-^0 \rightarrow F \rightarrow {\cal{H}}^-$\ .
\end{enumerate}
{}From Table 9,
which describes the initial and final states of all asymptotes
in the interior phase space (note the difference with the tables in the
previous section where the tables referred to a single equilibrium
point/line), it follows that
all models begin with a curvature singularity and
approach ${\cal{H}}^-$, with $0\leq\Sp\leq1/2$, at late times
(see Diagram 2a).
None of these endpoints correspond to curvature singularities,
although for $\Sp>1/5$ they are crushing singularities.

\def\Strut{\vrule width 0pt height 13pt depth 6pt{}}
\begin{table}[h]
\begin{center}
  \begin{tabular}{|c|c|c|c|c|c|c|c|c|c|}  \hline 
    \multicolumn{2}{|c|}{As.\,St.} & \Strut $\tilde{\mu}$ & $\tilde{\theta}$
 & $\tilde{\sigma}$
  & $\tilde{a}$ & ${\tilde{a}}_{;\mu}^{\mu}$ & $D_1$ & $D_2$ & $C$\\ \hline

  \Strut$F$ & ${\rm i}$ & $e^{-3\gamma\tau}$ & $e^{-3\gamma\tau /2}$ &
  $e^{(3\gamma-4)\tau /2}$ & $e^{(\gamma-1)\tau}$ &$e^{3(\gamma-4)\tau}$ &
  $ e^{\tau}$ & $e^{\tau}$ & $e^{-2\tau}$ \\ \hline
  \Strut${\cal{H}}^-$ & ${\rm f}$ & $e^{\Gamma_1\tau}$ & $e^{\Gamma_2\tau}$
  & $e^{\Gamma_2\tau}$ &
  $e^{\Gamma_2\tau}$ & $e^{2\Gamma_2\tau}$ & $e^{(2\Sigma_+-1)\tau}$
  & $e^{-\Gamma_1\tau/2}$ & $e^{\Gamma_1\tau}$\\ \hline \hline

  \Strut${\cal{H}}^-$ & ${\rm i}$ & $e^{\Gamma_1\tau}$ & $e^{\Gamma_2\tau}$
  & $e^{\Gamma_2\tau}$ &
  $e^{\Gamma_2\tau}$ & $e^{2\Gamma_2\tau}$ & $e^{(2\Sigma_+-1)\tau}$
  & $e^{-\Gamma_1\tau/2}$ & $e^{\Gamma_1\tau}$\\ \hline \hline
  \Strut${\cal{H}}^-$ & ${\rm f}$ & $e^{\Gamma_1\tau}$ & $e^{\Gamma_2\tau}$
  & $e^{\Gamma_2\tau}$ &
  $e^{\Gamma_2\tau}$ & $e^{2\Gamma_2\tau}$ & $e^{(2\Sigma_+-1)\tau}$
  & $e^{-\Gamma_1\tau/2}$ & $e^{\Gamma_1\tau}$\\ \hline \hline

 \Strut$K^0_+$ & ${\rm i}$ &  $e^{-3\gamma\tau}$ & $e^{-3\tau}$ &
 $e^{-3\tau}$ & $e^{(3\gamma-5)\tau}$ & $e^{2(3\gamma-5)\tau}$ &
 $e^{-\tau}$& $e^{2\tau}$ & $e^{-6\tau}$\\ \hline
\Strut${\cal{H}}^-$ & ${\rm f}$ & $e^{\Gamma_1\tau}$ & $e^{\Gamma_2\tau}$
  & $e^{\Gamma_2\tau}$ &
  $e^{\Gamma_2\tau}$ & $e^{2\Gamma_2\tau}$ & $e^{(2\Sigma_+-1)\tau}$
  & $e^{-\Gamma_1\tau/2}$ & $e^{\Gamma_1\tau}$\\ \hline
\end{tabular}
\end{center}
\caption{Initial and final points for the type $\mbox{}_1I$-models with
$1 \leq \gamma \leq 6/5$. The constants $\Gamma_{1,2}$ are given by
$\Gamma_1=-2(1+\Sigma_+)$ and $\Gamma_2=(5\Sigma_+-1)$.}
\label{tab:AIi}
\end{table}

\subsubsection{Models with $ 6/5 < \gamma < 2$}
The orbits are described by ${\cal{H}}^- \rightarrow
{\cal{H}}^-$, $F \rightarrow {\cal{H}}^-$, $K_+^0 \rightarrow {\cal{H}}^-$,
$K_+^0 \rightarrow M^{v_+}$ and $K_+^0 \rightarrow M^+$.
It follows from Table 10 that all models
start with a curvature singularity. None of the end-points
correspond to a curvature
singularity. Points on ${\cal{H}}^-$ with $\Sp>1/5$ and $M^+$ correspond to
crushing singularities. The remaining final points are not associated
with any singular behaviour at all.
The intermediate evolution of orbits can be approximated by the
following heteroclinic sequences:
\renewcommand{\labelenumi}{(\theenumi)}
\begin{enumerate}
 \item $K_+^0 \rightarrow K_+^+ \rightarrow M^+$
 \item $K_+^0 \rightarrow M^{v_+} \rightarrow M^+$
 \item $K_+^0 \rightarrow F \rightarrow {\cal{H}}^-$
 \item ${\cal{H}}^- \rightarrow K_-^0 \rightarrow F \rightarrow {\cal{H}}^-$\ .
\end{enumerate}

\def\Strut{\vrule width 0pt height 13pt depth 6pt{}}
\begin{table}[h]
\begin{center}
  \begin{tabular}{|c|c|c|c|c|c|c|c|c|c|}  \hline 
    \multicolumn{2}{|c|}{As.\,St.} & \Strut $\tilde{\mu}$ &
  $\tilde{\theta}$ & $\tilde{\sigma}$
  & $\tilde{a}$ & ${\tilde{a}}_{;\mu}^{\mu}$ & $D_1$ & $D_2$ & $C$\\ \hline

  \Strut$F$ & ${\rm i}$ & $e^{-3\gamma\tau}$ & $e^{-3\gamma\tau /2}$ &
  $e^{(3\gamma-4)\tau /2}$ & $e^{(\gamma-1)\tau}$ &$e^{3(\gamma-4)\tau}$ &
  $ e^{\tau}$ & $e^{\tau}$ & $e^{-2\tau}$ \\ \hline
  \Strut${\cal{H}}^-$ & ${\rm f}$ & $e^{\Gamma_1\tau}$ & $e^{\Gamma_2\tau}$
  & $e^{\Gamma_2\tau}$ &
  $e^{\Gamma_2\tau}$ & $e^{2\Gamma_2\tau}$ & $e^{(2\Sigma_+-1)\tau}$
  & $e^{-\Gamma_1\tau/2}$ & $e^{\Gamma_1\tau}$\\ \hline \hline

 \Strut$K^0_+$ & ${\rm i}$ &  $e^{-3\gamma\tau}$ & $e^{-3\tau}$ &
 $e^{-3\tau}$ & $e^{(3\gamma-5)\tau}$ & $e^{2(3\gamma-5)\tau}$ &
 $e^{-\tau}$& $e^{2\tau}$ & $e^{-6\tau}$\\ \hline
   \Strut$M^{v_+}$ & ${\rm f}$ & $e^{-\lambda^2\tau}$ & $\f$ & $\f$ & $\f$
   & $\f$ & $\f$&
   $e^{3\tau/2}$ & $e^{-\lambda^2\tau}$\\ \hline \hline

 \Strut$K^0_+$ & ${\rm i}$ &  $e^{-3\gamma\tau}$ & $e^{-3\tau}$ &
 $e^{-3\tau}$ & $e^{(3\gamma-5)\tau}$ & $e^{2(3\gamma-5)\tau}$ &
 $e^{-\tau}$& $e^{2\tau}$ & $e^{-6\tau}$\\ \hline
  \Strut${\cal{H}}^-$ & ${\rm f}$ & $e^{\Gamma_1\tau}$ & $e^{\Gamma_2\tau}$
  & $e^{\Gamma_2\tau}$ &
  $e^{\Gamma_2\tau}$ & $e^{2\Gamma_2\tau}$ & $e^{(2\Sigma_+-1)\tau}$
  & $e^{-\Gamma_1\tau/2}$ & $e^{\Gamma_1\tau}$\\ \hline \hline

  \Strut${\cal{H}}^-$ & ${\rm i}$ & $e^{\Gamma_1\tau}$ & $e^{\Gamma_2\tau}$
  & $e^{\Gamma_2\tau}$ &
  $e^{\Gamma_2\tau}$ & $e^{2\Gamma_2\tau}$ & $e^{(2\Sigma_+-1)\tau}$
  & $e^{-\Gamma_1\tau/2}$ & $e^{\Gamma_1\tau}$\\ \hline
  \Strut${\cal{H}}^-$ & ${\rm f}$ & $e^{\Gamma_1\tau}$ & $e^{\Gamma_2\tau}$
  & $e^{\Gamma_2\tau}$ &
  $e^{\Gamma_2\tau}$ & $e^{2\Gamma_2\tau}$ & $e^{(2\Sigma_+-1)\tau}$
  & $e^{-\Gamma_1\tau/2}$ & $e^{\Gamma_1\tau}$\\ \hline \hline

 \Strut$K^0_+$ & ${\rm i}$ &  $e^{-3\gamma\tau}$ & $e^{-3\tau}$ &
 $e^{-3\tau}$ & $e^{(3\gamma-5)\tau}$ & $e^{2(3\gamma-5)\tau}$ &
 $e^{-\tau}$& $e^{2\tau}$ & $e^{-6\tau}$\\ \hline
 \Strut$M^+$ & ${\rm f}$ & $e^{\Gamma_4\tau}$ &
   $e^{\Gamma_3\tau}$ & $e^{\Gamma_3\tau}$ &
   $e^{\Gamma_3\tau}$ &
   $e^{2\Gamma_3\tau}$ & $e^{3a\tau/(3a+2f)}$& $e^{\Gamma_5}$ &
   $e^{-6\tau}$ \\ \hline

 \Strut$K^0_+, \gamma=4/3$ & ${\rm i}$ &  $e^{-3\gamma\tau}$ & $e^{-3\tau}$ &
 $e^{-3\tau}$ & $e^{(3\gamma-5)\tau}$ & $e^{2(3\gamma-5)\tau}$ &
 $e^{-\tau}$& $e^{2\tau}$ & $e^{-6\tau}$\\ \hline
 \Strut$M^+,\gamma=4/3$ & ${\rm f}$ & $e^{\Gamma_4\tau}$ &
   $e^{\Gamma_3\tau}$ & $e^{-\tau/2}$ &
   $e^{\Gamma_3\tau}$ &
   $e^{2\Gamma_3\tau}$ & $e^{3a\tau/(3a+2f)}$& $e^{\Gamma_5}$ &
   $e^{-6\tau}$ \\ \hline
\end{tabular}
\end{center}
\caption{Initial and final points for the type $\mbox{}_1I$-models with
$6/5 < \gamma <2$. The constants $\Gamma_\mu$, $\mu = 1-5$, are given by
$\Gamma_1=-2(1+\Sigma_+)$,
$\Gamma_2=(5\Sigma_+-1)$, $\Gamma_3=3(5\gamma-6)/\left(2(2-\gamma)\right)$,
$\Gamma_4=3(3\gamma-2)\tau/(2-\gamma)$ and $\Gamma_5=3(a+f)\tau/(3a+2f)$.
Here and below $\f$ is an abbreviation for finite constant.}
\label{tab:AIii}
\end{table}

\begin{figure}[h]
    \begin{center}
      \leavevmode

      \vspace{10.0cm}

    \end{center}
      \caption{The reduced phase space of the type $\mbox{}_1I$ models
with (a) $\gamma \leq 6/5$ and (b) $6/5 < \gamma < 2$.}
  \label{fig:dia2}
\end{figure}

\subsection{Some comments on LRS spatially homogeneous type V models}

The SH LRS type V models have been extensively discussed in the literature
\cite{hew,coel,shiv,col}. However, we will here make some remarks about
the existence of monotonic functions. The SH type V models differ from
the SSS type $_f$V and $_1$I models
in that there exists an exact solution corresponding to
an invariant submanifold described by
$v=0, \Sp = 0$ (this is just the open Friedmann-Robertson-Walker (FRW)
model). This separatrix divides the reduced 2-dimensional phase space
of the SH type V models into
two regions. For the type $_f$V and $_1$I
models one also has a separatrix leaving the equilibrium
point $F$. However, at what equilibrium point this separatrix ends up at
depends on $\gamma$ and $f$, and so far one has not been able to find
the corresponding exact solution.
The simple division of the phase space created by the FRW separatrix
in the SH type V models allows one to find monotonic
functions valid in the $v > 0$ and $v < 0$ regions respectively.

The conditions $A, \Omega_n > 0$ (for the interior matter
part of the type V phase space) and the constraint, lead to $v \Sp < 0$.
Let us first consider $v < 0$. Then $\Sp$ ($A$) is a monotonically
decreasing (increasing) function for $1 \leq \gamma < 2$. The variable
$v$ is monotonically decreasing when $4/3 \leq \gamma < 2$.
If $v > 0$ then $v$ is monotonically decreasing if
$1 \leq \gamma \leq 4/3$.
Thus one can obtain considerable qualitative dynamical information
because of the simple separatrix structure.
Unfortunately this structure is not available for the SSS models.


\subsection{Models with radiation, $\gamma=4/3$}
The general structure of the type $_f$V models is complicated by the number
of parameters. To avoid being lost in details we here therefore restrict
ourselves to the physically interesting radiation case, $\gamma = 4/3$.
For $\gamma = 4/3$ we obtain a number of
bifurcations for different values of the parameters $a$ and $f$.

\subsubsection{Type $\mbox{}_f$V models with $3a+f<0$}
The phase space has the same structure as
for the type $\mbox{}_1$I models with $\gamma=4/3$.
It follows from Diagram $3a$ that the
initial and final end points of the orbits are described by
${\cal{H}}^- \rightarrow
{\cal{H}}^-$, $F \rightarrow {\cal{H}}^-$, $K_+^0 \rightarrow {\cal{H}}^-$,
$K_+^0 \rightarrow M_{v_+}$ and $K_+^0 \rightarrow M^+$. From
Table 11 it is seen that all models start with a curvature
singularity and that the point $M^+$ corresponds to a crushing
singularity. If the final points on ${\cal{H}}^-$ are crushing singularities
or not depends on which of the inequalities given in section 4 the
final value of $\Sp$ satisfies. It is clear though, that the line has points
corresponding to crushing singularities or no singularities at all.
We have the following heteroclinic sequences:
\renewcommand{\labelenumi}{(\theenumi)}
\begin{enumerate}
 \item $K_+^0 \rightarrow K_+^+ \rightarrow M^+$
 \item $K_+^0 \rightarrow M^{v_+} \rightarrow M^+$
 \item $K_+^0 \rightarrow F \rightarrow {\cal{H}}^-$
 \item ${\cal{H}}^- \rightarrow K_-^0 \rightarrow
 F  \rightarrow {\cal{H}}^-$\ .
\end{enumerate}


\def\Strut{\vrule width 0pt height 13pt depth 6pt{}}
\begin{table}[h]
\begin{center}
  \begin{tabular}{|c|c|c|c|c|c|c|c|c|c|}  \hline 
    \multicolumn{2}{|c|}{As.\,St.} & \Strut $\tilde{\mu}$ &
  $\tilde{\theta}$ & $\tilde{\sigma}$
  & $\tilde{a}$ & ${\tilde{a}}_{;\mu}^{\mu}$ & $D_1$ & $D_2$ & $C$\\ \hline

  \Strut$F$ & ${\rm i}$ & $e^{-4\tau}$ & $e^{-2\tau}$ &
  $\f$ & $e^{\tau/3}$ &$e^{-8\tau}$ &
  $ e^{\tau}$ & $e^{\tau}$ & $e^{-2\tau}$ \\ \hline
  \Strut${\cal{H}}^-$ & ${\rm f}$ & $e^{\Gamma_1\tau}$ & $e^{\Gamma_2\tau}$
  & $e^{\Gamma_2\tau}$ &
  $e^{\Gamma_2\tau}$ & $e^{2\Gamma_2\tau}$ & $e^{(2\Sigma_+-1)\tau}$
  & $e^{-\Gamma_1\tau/2}$ & $e^{\Gamma_1\tau}$\\ \hline \hline

 \Strut$K^0_+$ & ${\rm i}$ &  $e^{-4\tau}$ & $e^{-4\tau}$ &
 $e^{-3\tau}$ & $e^{-\tau}$ & $e^{-2\tau}$ &
 $e^{-\tau}$& $e^{2\tau}$ & $e^{-6\tau}$\\ \hline
   \Strut$M^{v_+}$ & ${\rm f}$ & $e^{-\lambda^2\tau}$ & $\f$ & $\f$
  & $\f$ & $\f$ & $\f$&
   $e^{3\tau/2}$ & $e^{-\lambda^2\tau}$\\ \hline \hline

 \Strut$K^0_+$ & ${\rm i}$ &  $e^{-4\tau}$ & $e^{-3\tau}$ &
 $e^{-3\tau}$ & $e^{-\tau}$ & $e^{-2\tau}$ &
 $e^{-\tau}$& $e^{2\tau}$ & $e^{-6\tau}$\\ \hline
  \Strut${\cal{H}}^-$ & ${\rm f}$ & $e^{\Gamma_1\tau}$ & $e^{\Gamma_2\tau}$
  & $e^{\Gamma_2\tau}$ &
  $e^{\Gamma_2\tau}$ & $e^{2\Gamma_2\tau}$ & $e^{(2\Sigma_+-1)\tau}$
  & $e^{-\Gamma_1\tau/2}$ & $e^{\Gamma_1\tau}$\\ \hline \hline

  \Strut${\cal{H}}^-$ & ${\rm i}$ & $e^{\Gamma_1\tau}$ & $e^{\Gamma_2\tau}$
  & $e^{\Gamma_2\tau}$ &
  $e^{\Gamma_2\tau}$ & $e^{2\Gamma_2\tau}$ & $e^{(2\Sigma_+-1)\tau}$
  & $e^{-\Gamma_1\tau/2}$ & $e^{\Gamma_1\tau}$\\ \hline
  \Strut${\cal{H}}^-$ & ${\rm f}$ & $e^{\Gamma_1\tau}$ & $e^{\Gamma_2\tau}$
  & $e^{\Gamma_2\tau}$ &
  $e^{\Gamma_2\tau}$ & $e^{2\Gamma_2\tau}$ & $e^{(2\Sigma_+-1)\tau}$
  & $e^{-\Gamma_1\tau/2}$ & $e^{\Gamma_1\tau}$\\ \hline \hline

 \Strut$K^0_+$ & ${\rm i}$ &  $e^{-4\tau}$ & $e^{-3\tau}$ &
 $e^{-3\tau}$ & $e^{-\tau}$ & $e^{-2\tau}$ &
 $e^{-\tau}$& $e^{2\tau}$ & $e^{-6\tau}$\\ \hline
 \Strut$M^+$ & ${\rm f}$ & $e^{\Gamma_4\tau}$ &
   $e^{\Gamma_3\tau}$ & $e^{\Gamma_3\tau}$ &
   $e^{\Gamma_3\tau}$ &
   $e^{2\Gamma_3\tau}$ & $e^{3a\tau/(3a+2f)}$& $e^{3(a+f)\tau/(3a+2f)}$ &
   $e^{-6\tau}$ \\ \hline
\end{tabular}
\end{center}
\caption{Initial and final points for the type $\mbox{}_fV$-models with
$\gamma=4/3$ and $(3a+f)<0$. The constants $\Gamma_{1,2,3,4}$ are given by
$\Gamma_1=-2f(1+\Sigma_+)/(a+f)$,
$\Gamma_2=((5\Sigma_+-1)f-2a(1-2\Sigma_+))/(a+f)$,
$\Gamma_3=-\frac{2a+f}{3a+2f}$ and $\Gamma_4=-6(4a+3f)/(3a+2f)$.
}
\label{tab:43a}
\end{table}


\subsubsection{Type $\mbox{}_f$V models with $a+f>0$ and $f<0$}
The orbits are described by
${\cal{H}}^+ \rightarrow M^-$, $F \rightarrow M^-$, and
$K^0_+ \rightarrow M^-$ (see Diagram $3b$). It follows from Table 12
that the orbits starting from ${\cal{H}}^+$ do not
have an initial curvature singularity.
The other two types of orbits have an initial curvature singularity.
The heteroclinic sequences are given by:
\renewcommand{\labelenumi}{(\theenumi)}
\begin{enumerate}
 \item ${\cal{H}}^+ \rightarrow M^-$
 \item ${\cal{H}}^+ \rightarrow K_-^0 \rightarrow F \rightarrow M^-$
 \item $K_+^0 \rightarrow F \rightarrow M^-$
 \item $K_+^0 \rightarrow K_-^0 \rightarrow M^-$\ .
\end{enumerate}
As seen in Diagram $3b$, {\it all} orbits approach $M^-$ for late times, and
as can be seen from Table 12,
this point corresponds to no singular behaviour at all.

\def\Strut{\vrule width 0pt height 13pt depth 6pt{}}
\begin{table}[h]
\begin{center}
  \begin{tabular}{|c|c|c|c|c|c|c|c|c|c|}  \hline 
    \multicolumn{2}{|c|}{As.\,St.} & \Strut $\tilde{\mu}$ &
  $\tilde{\theta}$ & $\tilde{\sigma}$
  & $\tilde{a}$ & ${\tilde{a}}_{;\mu}^{\mu}$ & $D_1$ & $D_2$ & $C$\\ \hline

  \Strut$F$ & ${\rm i}$ & $e^{-4\tau}$ & $e^{-2\tau}$ &
  $\f$ & $e^{\tau/3}$ &$e^{-8\tau}$ &
  $ e^{\tau}$ & $e^{\tau}$ & $e^{-2\tau}$ \\ \hline
 \Strut$M^-$ & ${\rm f}$ & $e^{\Gamma_4\tau}$ &
   $e^{\Gamma_3\tau}$ & $e^{\Gamma_3\tau}$ &
   $e^{\Gamma_3\tau}$ &
   $e^{2\Gamma_3\tau}$ & $e^{3a\tau/(3a+2f)}$& $e^{3(a+f)\tau/(3a+2f)}$ &
   $e^{-6\tau}$ \\ \hline \hline

  \Strut$K^0_+$ & ${\rm i}$ &  $e^{-4\tau}$ & $e^{-4\tau}$ &
 $e^{-3\tau}$ & $e^{-\tau}$ & $e^{-2\tau}$ &
 $e^{-\tau}$& $e^{2\tau}$ & $e^{-6\tau}$\\ \hline
 \Strut$M^-$ & ${\rm f}$ & $e^{\Gamma_4\tau}$ &
   $e^{\Gamma_3\tau}$ & $e^{\Gamma_3\tau}$ &
   $e^{\Gamma_3\tau}$ &
   $e^{2\Gamma_3\tau}$ & $e^{3a\tau/(3a+2f)}$& $e^{3(a+f)\tau/(3a+2f)}$ &
   $e^{-6\tau}$ \\ \hline \hline

  \Strut${\cal{H}}^+$ & ${\rm i}$ & $e^{\Gamma_1\tau}$ & $e^{\Gamma_2\tau}$
  & $e^{\Gamma_2\tau}$ &
  $e^{\Gamma_2\tau}$ & $e^{2\Gamma_2\tau}$ & $e^{(2\Sigma_+-1)\tau}$
  & $e^{-\Gamma_1\tau/2}$ & $e^{\Gamma_1\tau}$\\ \hline
 \Strut$M^-$ & ${\rm f}$ & $e^{\Gamma_4\tau}$ &
   $e^{\Gamma_3\tau}$ & $e^{\Gamma_3\tau}$ &
   $e^{\Gamma_3\tau}$ &
   $e^{2\Gamma_3\tau}$ & $e^{3a\tau/(3a+2f)}$& $e^{3(a+f)\tau/(3a+2f)}$ &
   $e^{-6\tau}$ \\ \hline
\end{tabular}
\end{center}
\caption{Initial and final points for the type $\mbox{}_fV$-models with
$\gamma=4/3$ and $(a+f)>, f<0$. The constants $\Gamma_{1,2,3,4}$ are given by
$\Gamma_1=-2f(1+\Sigma_+)
/(a+f)$, $\Gamma_2=((5\Sigma_+-1)f-2a(1-2\Sigma_+))/(a+f)$,
$\Gamma_3=-\frac{2a+f}{3a+2f}$ and $\Gamma_4=-6(4a+3f)/(3a+2f)$.}
\label{tab:43b}
\end{table}


\subsubsection{Type V models}
For $f=0$, $v=0$ is an invariant submanifold, characterized by the separatrix
$F \rightarrow M^0$. The other orbits are characterized by
${\cal{H}}^+ \rightarrow
M^0$ and $K^0_+ \rightarrow M^-$ (see Diagram $3c$).
Orbits starting from the points $K^0_+$ and $F$ have an initial
curvature singularity while orbits starting from ${\cal{H}}^+$
do not.
The point $M^0$, which only exists for these models,
has a zero eigenvalue for $\gamma=4/3$ and needs special
treatment. We therefore refrain from giving a table of initial and final
asymptotic expressions and instead refer to \cite{coel}.
It follows from Table 7 that $M^-$ is associated with no singular
behaviour at all while Table 8 shows that the line ${\cal{H}^+}$
corresponds to a crushing singularity.
The intermediate evolution of orbits
can be approximated by the following heteroclinic sequences:
\renewcommand{\labelenumi}{(\theenumi)}
\begin{enumerate}
 \item ${\cal{H}}^+ \rightarrow K_-^0 \rightarrow F \rightarrow M^0$
 \item $K_+^0 \rightarrow F \rightarrow M^0 \rightarrow M^-$
 \item $K_+^0 \rightarrow K_-^0 \rightarrow M^-$\ .
\end{enumerate}

\subsubsection{Type $\mbox{}_f$V models with $a+f>0$ and $0<f<a$}
In this case the
SH type V point $M^0$ splits into the two points
$M^{v_\pm}$. These points move apart as $f$ increases.
When $f$ approaches $a$, $M^{v_+}$ approaches
the point ${\bar{M}}^+$.
The initial and final stages of the orbits are given by
${\cal{H}}^+ \rightarrow M^{v_+}$, $F \rightarrow M^{v_+}$,
$K^0_+ \rightarrow M^{v_+}$, $K^0_+ \rightarrow M^{v_-}$ and $K^0_+
\rightarrow M^-$ (see Diagram $3d$). It follows from Table 13
that all orbits have an initial
curvature singularity. Solutions ending in $M^-$ have a final crushing
singularity.
The intermediate evolution of orbits
can be approximated by the following heteroclinic sequences:
\renewcommand{\labelenumi}{(\theenumi)}
\begin{enumerate}
 \item ${\cal{H}}^+ \rightarrow K_-^0 \rightarrow F \rightarrow M^{v_+}$
 \item $K_+^0 \rightarrow F \rightarrow M^{v_+}$
 \item $K_+^0 \rightarrow M^{v_-} \rightarrow M^{v_+}$
 \item $K_+^0 \rightarrow M^{v_-} \rightarrow M^-$
 \item $K_+^0 \rightarrow K_+^- \rightarrow M^-$\ .
\end{enumerate}

\def\Strut{\vrule width 0pt height 13pt depth 6pt{}}
\begin{table}[h]
\begin{center}
  \begin{tabular}{|c|c|c|c|c|c|c|c|c|c|}  \hline 
    \multicolumn{2}{|c|}{As.\,St.} & \Strut $\tilde{\mu}$ &
  $\tilde{\theta}$ & $\tilde{\sigma}$
  & $\tilde{a}$ & ${\tilde{a}}_{;\mu}^{\mu}$ & $D_1$ & $D_2$ & $C$\\ \hline

  \Strut$F$ & ${\rm i}$ & $e^{-4\tau}$ & $e^{-2\tau}$ &
  $\f$ & $e^{\tau/3}$ &$e^{-8\tau}$ &
  $ e^{\tau}$ & $e^{\tau}$ & $e^{-2\tau}$ \\ \hline
   \Strut$M^{v_+}$ & ${\rm f}$ & $e^{-\lambda^2\tau}$ & $\f$ & $\f$
& $\f$ & $\f$ & $\f$&
   $e^{3\tau/2}$ & $e^{-\lambda^2\tau}$\\ \hline \hline

 \Strut$K^0_+$ & ${\rm i}$ &  $e^{-4\tau}$ & $e^{-4\tau}$ &
 $e^{-3\tau}$ & $e^{-\tau}$ & $e^{-2\tau}$ &
 $e^{-\tau}$& $e^{2\tau}$ & $e^{-6\tau}$\\ \hline
   \Strut$M^{v_+}$ & ${\rm f}$ & $e^{-\lambda^2\tau}$ & $\f$ & $\f$ & $\f$
& $\f$ & $\f$&
   $e^{3\tau/2}$ & $e^{-\lambda^2\tau}$\\ \hline \hline

 \Strut$K^0_+$ & ${\rm i}$ &  $e^{-4\tau}$ & $e^{-4\tau}$ &
 $e^{-3\tau}$ & $e^{-\tau}$ & $e^{-2\tau}$ &
 $e^{-\tau}$& $e^{2\tau}$ & $e^{-6\tau}$\\ \hline
   \Strut$M^{v_-}$ & ${\rm f}$ & $e^{-\lambda_1^2\tau}$ & $\f$ & $\f$ & $\f$
 & $\f$ & $\f$&
   $e^{3\tau/2}$ & $e^{-\lambda_1^2\tau}$\\ \hline \hline

  \Strut${\cal{H}}^+$ & ${\rm i}$ & $e^{\Gamma_1\tau}$ & $e^{\Gamma_2\tau}$
  & $e^{\Gamma_2\tau}$ &
  $e^{\Gamma_2\tau}$ & $e^{2\Gamma_2\tau}$ & $e^{(2\Sigma_+-1)\tau}$
  & $e^{-\Gamma_1\tau/2}$ & $e^{\Gamma_1\tau}$\\ \hline \hline
   \Strut$M^{v_+}$ & ${\rm f}$ & $e^{-\lambda_2^2\tau}$ & $\f$ & $\f$ & $\f$
& $\f$ & $\f$&
   $e^{3\tau/2}$ & $e^{-\lambda_2^2\tau}$\\ \hline \hline

 \Strut$K^0_+$ & ${\rm i}$ &  $e^{-4\tau}$ & $e^{-3\tau}$ &
 $e^{-3\tau}$ & $e^{-\tau}$ & $e^{-2\tau}$ &
 $e^{-\tau}$& $e^{2\tau}$ & $e^{-6\tau}$\\ \hline
 \Strut$M^-$ & ${\rm f}$ & $e^{\Gamma_4\tau}$ &
   $e^{\Gamma_3\tau}$ & $e^{\Gamma_3\tau}$ &
   $e^{\Gamma_3\tau}$ &
   $e^{2\Gamma_3\tau}$ & $e^{3a\tau/(3a+2f)}$& $e^{3(a+f)\tau/(3a+2f)}$ &
   $e^{-6\tau}$ \\ \hline
\end{tabular}
\end{center}
\caption{Initial and final points for the type $\mbox{}_fV$-models with
$\gamma=4/3$ and $(a+f)>0, 0 < f < a$. The constants $\Gamma_{1,2,3,4}$ are
given by $\Gamma_1=
-2f(1+\Sigma_+)/(a+f)$, $\Gamma_2=((5\Sigma_+-1)f-2a(1-2\Sigma_+))/(a+f)$,
$\Gamma_3=-\frac{2a+f}{3a+2f}$ and $\Gamma_4=-6(4a+3f)/(3a+2f)$.
The constants $\lambda_{1,2}^2$ have long expressions which will not be given.}
\label{tab:43d}
\end{table}


\subsubsection{Type $\mbox{}_f$V models with $a+f>0$ and $f \geq a$}
The asymptotic properties of the orbits are described by
${\cal{H}}^+ \rightarrow
{\cal{H}}^+$, $F \rightarrow {\cal{H}}^+$, $K_+^0 \rightarrow {\cal{H}}^+$,
$K_+^0 \rightarrow M_{v_-}$ and $K_+^0 \rightarrow M^-$ (see Diagram $3e$).
{}From Table 14 it is seen that all models have an initial curvature
singularity. From Table 14 it also follows that orbits ending in
the point $M^-$ have a crushing singularity.
A solution ending on the line ${\cal{H}}^+$ may have a final crushing
singularity or no singularity at all. The existence of a final
crushing singularity depends
on which of the previously given inequalities the final value of $\Sp$
satisfies (see section 4).
We have the following heteroclinic sequences:
\renewcommand{\labelenumi}{(\theenumi)}
\begin{enumerate}
 \item $K_+^0 \rightarrow K_+^- \rightarrow M^-$
 \item $K_+^0 \rightarrow M^{v_-} \rightarrow M^-$
 \item $K_+^0 \rightarrow F \rightarrow {\cal{H}}^+$
 \item $K_+^0 \rightarrow M^{v_-} \rightarrow {\cal{H}}^+$
 \item ${\cal{H}}^+ \rightarrow K_-^0 \rightarrow
 F \rightarrow {\cal{H}}^+$\ .
\end{enumerate}

\def\Strut{\vrule width 0pt height 13pt depth 6pt{}}
\begin{table}[ht]
\begin{center}
  \begin{tabular}{|c|c|c|c|c|c|c|c|c|c|}  \hline 
    \multicolumn{2}{|c|}{As.\,St.} & \Strut $\tilde{\mu}$ &
  $\tilde{\theta}$ & $\tilde{\sigma}$
  & $\tilde{a}$ & ${\tilde{a}}_{;\mu}^{\mu}$ & $D_1$ & $D_2$ & $C$\\ \hline

  \Strut$F$ & ${\rm i}$ & $e^{-4\tau}$ & $e^{-2\tau}$ &
  $\f$ & $e^{\tau/3}$ &$e^{-8\tau}$ &
  $ e^{\tau}$ & $e^{\tau}$ & $e^{-2\tau}$ \\ \hline
  \Strut${\cal{H}}^+$ & ${\rm f}$ & $e^{\Gamma_1\tau}$ & $e^{\Gamma_2\tau}$
  & $e^{\Gamma_2\tau}$ &
  $e^{\Gamma_2\tau}$ & $e^{2\Gamma_2\tau}$ & $e^{(2\Sigma_+-1)\tau}$
  & $e^{-\Gamma_1\tau/2}$ & $e^{\Gamma_1\tau}$\\ \hline \hline

 \Strut$K^0_+$ & ${\rm i}$ &  $e^{-4\tau}$ & $e^{-4\tau}$ &
 $e^{-3\tau}$ & $e^{-\tau}$ & $e^{-2\tau}$ &
 $e^{-\tau}$& $e^{2\tau}$ & $e^{-6\tau}$\\ \hline
   \Strut$M^{v_+}$ & ${\rm f}$ & $e^{-\lambda^2\tau}$ & $\f$ & $\f$ & $\f$
& $\f$ & $\f$&
   $e^{3\tau/2}$ & $e^{-\lambda^2\tau}$\\ \hline \hline

 \Strut$K^0_+$ & ${\rm i}$ &  $e^{-4\tau}$ & $e^{-3\tau}$ &
 $e^{-3\tau}$ & $e^{-\tau}$ & $e^{-2\tau}$ &
 $e^{-\tau}$& $e^{2\tau}$ & $e^{-6\tau}$\\ \hline
  \Strut${\cal{H}}^+$ & ${\rm f}$ & $e^{\Gamma_1\tau}$ & $e^{\Gamma_2\tau}$
  & $e^{\Gamma_2\tau}$ &
  $e^{\Gamma_2\tau}$ & $e^{2\Gamma_2\tau}$ & $e^{(2\Sigma_+-1)\tau}$
  & $e^{-\Gamma_1\tau/2}$ & $e^{\Gamma_1\tau}$\\ \hline \hline

  \Strut${\cal{H}}^+$ & ${\rm i}$ & $e^{\Gamma_1\tau}$ & $e^{\Gamma_2\tau}$
  & $e^{\Gamma_2\tau}$ &
  $e^{\Gamma_2\tau}$ & $e^{2\Gamma_2\tau}$ & $e^{(2\Sigma_+-1)\tau}$
  & $e^{-\Gamma_1\tau/2}$ & $e^{\Gamma_1\tau}$\\ \hline \hline
  \Strut${\cal{H}}^+$ & ${\rm f}$ & $e^{\Gamma_1\tau}$ & $e^{\Gamma_2\tau}$
  & $e^{\Gamma_2\tau}$ &
  $e^{\Gamma_2\tau}$ & $e^{2\Gamma_2\tau}$ & $e^{(2\Sigma_+-1)\tau}$
  & $e^{-\Gamma_1\tau/2}$ & $e^{\Gamma_1\tau}$\\ \hline \hline

 \Strut$K^0_+$ & ${\rm i}$ &  $e^{-4\tau}$ & $e^{-3\tau}$ &
 $e^{-3\tau}$ & $e^{-\tau}$ & $e^{-2\tau}$ &
 $e^{-\tau}$& $e^{2\tau}$ & $e^{-6\tau}$\\ \hline
 \Strut$M^-$ & ${\rm f}$ & $e^{\Gamma_4\tau}$ &
   $e^{\Gamma_3\tau}$ & $e^{\Gamma_3\tau}$ &
   $e^{\Gamma_3\tau}$ &
   $e^{2\Gamma_3\tau}$ & $e^{3a\tau/(3a+2f)}$& $e^{3(a+f)\tau/(3a+2f)}$ &
   $e^{-6\tau}$ \\ \hline
\end{tabular}
\end{center}
\caption{Initial and final points for the type $\mbox{}_fV$-models with
$\gamma=4/3$ and $(a+f)>0, f \geq a$. The constants $\Gamma_{1,2,3,4}$ are
given by $\Gamma_1=
-2f(1+\Sigma_+)
/(a+f)$, $\Gamma_2=((5\Sigma_+-1)f-2a(1-2\Sigma_+))/(a+f)$,
$\Gamma_3=-\frac{2a+f}{3a+2f}$ and $\Gamma_4=-6(4a+3f)/(3a+2f)$.}
\label{tab:43e}
\end{table}

\subsection{Some remarks}
All models starting from points with non-extreme tilt start from the points
$F$ or $K_+^0$ and are characterized by curvature singularities.
Models belonging to the class $a+f>0$ and
$f\leq 0$, which initially have extreme tilt, may be extended so that
they come from the TSS region. The
remaining models on the other hand have an initial curvature singularity.
A solution which ends at a point corresponding to extreme tilt
can be extended to
the TSS region since these points do not correspond to a curvature
singularity (they only correspond to a crushing singularity or
no singularity at all).
There are no interior equilibrium points in any of the $\theta$-dominated type
$_f$V models, and hence no limit
cycles. Furthermore, consideration of the orbits on the boundary leads to the
conclusion that there are no heteroclinic cycles either.

\vfill

\newpage


\begin{figure}[h]

    \begin{center}
      \leavevmode

      \vspace{18.0cm}

    \end{center}
  \caption{The reduced phase space of the $\mbox{}_fV$ models with
    (a) $3a+f<0$, (b) $a+f>0$ and $f<0$, (c) $a>0$ and $f=0$, (d)
  $a+f>0$ and $0<f<a$ and (e) $a+f>0$ and $a \geq 0$.}
  \label{fig:dia3}
\end{figure}

\clearpage

\newpage

\section{Discussion}

Note that we do not obtain {\it any\/} final conformal
singularities (i.e., curvature singularities for which only the Weyl
tensor blows up) at the Cauchy horizon since $C \rightarrow 0$ whenever
$\mu \rightarrow 0, const$. This is in contrast to
\cite{coel} where it is claimed that one has a conformal singularity at
the horizon for the SH type V models with $4/3 < \gamma <2$ (see the lower
part of figure 7 in \cite{coel}). We only obtain a
crushing singularity in this case
and thus one should be able to extend the space time
beyond the Cauchy horizon. This result
is supported by our asymptotic analysis (done with help of the computer
algebra systems REDUCE and the SHEEP package CLASSI)
as well as numerical calculations
for the orbits throughout their entire evolution
Thus our result
supports conjecture 2 in \cite{elking} (p. 147) which states that
matter flow lines do not end at a conformal curvature singularity.

Note that for the type
$_f$V models it is {\it only} in the SH type V model one has
$\mu = const.\neq 0$ on the horizon when
$1 \leq\gamma < 2$ (this follows from Table 8 where we have
$\mu \propto e^{-2f(1+\Sigma_+)/(a+f)}$).
Collins has shown that only stiff
perfect fluids ($\gamma = 2$) allow
$\mu=0$ on the horizon \cite{col} in the SH case
(this implies that there cannot be any
fluid flow over the horizon).
This lead him to the conclusion
that this behaviour probably was associated with a rather esoteric equation
of state. However, for the self-similar models presently considered
one always has $\mu = 0$ or $\mu = \infty$ at
the horizon. When the SH type V models are seen as a special case of the
present class of models it is rather the case $\mu=const.\neq 0$ which
is exceptional. Moreover, the existence of whimper singularities
is intimately connected with $\mu=const.\neq 0$
(see \cite{elking} p. 136). Hence we believe
that no whimper singularities occur in the present self-similar models
(a proof would require a more subtle comparison of the
decrease in $\mu$ and the increase associated with the growth of the affine
parameter; see \cite{elking}).
The present analysis (exemplified by the result
$\mu \propto e^{-2f(1+\Sigma_+)/(a+f)}$)
thus reinforce the impression that whimper singularities are highly special.

\appendix

\section{Properties of the fluid congruence and Petrov type conditions}

The fluid expansion is given by
\be\eqalign{
  \tilde{\theta} &= \frac{e^{fx}}{(1-v^2)^{3/2}}\left\{v\dot{v}+(1-v^2)\left(
\theta-(2a+3f)B_1v\right)\right\}= \cr
  &= \frac{e^{fx}}{(1-v^2)^{3/2}}\left\{\frac{vv^{\prime}}{3}+(1-v^2)\left(
1-(2a+3f)Av\right)\right\}\theta \ .
}
\ee

The fluid shear scalar is
\be\eqalign{
  \tilde{\sigma} &= \sqrt{\fr{1}{2}\tilde{\sigma}_{ab}\tilde{\sigma}^{ab}}
  =\frac{e^{fx}}{\sqrt{3}(1-v^2)^{3/2}}\left| v\dot{v}-\left( 1-v^2 \right)
  (\sigma_+-aB_1v) \right|= \cr
  &= \frac{e^{fx}}{\sqrt{3}(1-v^2)^{3/2}}\left|
      \left\{\frac{vv^{\prime}}{3}-(1-v^2)
      \left(\Sigma_+-aAv\right)\right\}\theta\right|\ .
}
\ee

The fluid acceleration scalar is
\be\eqalign{
  \tilde{a} &= \sqrt{\tilde{a}^\mu\tilde{a}_\mu}=\frac{e^{fx}}{3(1-v^2)^{3/2}}
  \left|3\dot{v}+(1-v^2)\left(\theta v - 2\sigma_+ v-3fB_1 \right)\right|= \cr
  &= \frac{e^{fx}}{3(1-v^2)^{3/2}}\left|\left\{v^{\prime}+(1-v^2)(v-2\Sigma_+v-
  3fA)\right\}\theta\right|\ .
}
\ee

The divergence of the acceleration vector is given by
\be\eqalign{
  {\tilde{u}}^a{}_{;a} &=\frac{e^{2fx}}{3\left(1-v^2\right)^3}(
  3v(1-v^2)\ddot{v} - 3(1+3v^2){\dot{v}}^2 \cr
  &+ (1-v^2)(5v\theta-4v\sigma_+-3fv^2B_1-3(2a+3f)B_1)\dot{v}\cr
  & - (1-v^2)(3fv\dot{B}_1+(2\dot{\sigma}_+-\dot{\theta}
  -\theta^2+2\theta\sigma_+)v^2 - 2B_1(a+f)(2v\sigma_+-3fB_1))) \cr
  &=\frac{e^{2fx}}{3(1-v^2)^3}\left( \frac{1}{3}v\left(1-v^2\right)
  v^{\prime\prime}
  +\frac{1}{3}\left(1+3v^2\right)\left(v^{\prime}\right)^2 \right. \cr
  &-\frac{1}{3}\left(1-v^2\right)\left(3A_1(2a+3f+fv^2)+4v\Sigma_+ -5v
  \right)v^{\prime} \cr
  &+\frac{1}{3}(1+q)\left(1-v^2\right)\left(\left(1-v^2\right)\left(
  3fvA_1+2v^2\Sigma_+-v\right) -vv^{\prime}\right) \cr
  &-\left(1-v^2\right)\left(2A_1(a+f)-v\right)\left(3fA_1+2v\Sigma_+-v\right)
  \left. +\frac{1}{3}\left(3fvA_1^{\prime}+2v^2\Sp^{\prime}\right)
  \right)\theta^2
}\ee

The Weyl scalar is
\be\eqalign{
   C &=\sqrt{C_{abcd}C^{abcd}} =
   \frac{2e^{2fx}}{3\sqrt{3}}\left|-3\dot{\sigma}_+
    +3kB_2^2-\theta\sigma_++2\sigma_+^2\right|= \cr
   &\quad = \frac{2e^{2fx}}{3\sqrt{3}}\left|\left\{\Sigma_+^{\prime}+
   (2\Sigma_++q)\Sigma_++3kK\right\}\theta^2\right|\ .
}
\ee

Note that the magnetic part of the Weyl tensor is identically zero for all
models. The space-time is of  Petrov type 0 if
\be
  -3\dot{\sigma}_+-\sigma_+\theta+2\sigma_+^2+3kB_2^2 = 0\ ,
\ee
which can be written as
\be
  (\Sigma_+^{\prime}+(2\Sigma_++q)\Sigma_+ +3kK)\theta^2 = 0\ ,
\ee
otherwise it is of type D.

\section{Relation to the fluid approach}

The fluid approach, where one uses a coordinate system adapted to the
fluid velocity, is more common than the ``homothetic'' approach used in this
article.
The line element, of the presently considered LRS models, takes
the following form in the fluid approach \cite{ship}:
\be
   d{\tilde s}^2 = - e^{\Psi(\lambda)}dT^2 + e^{\Lambda(\lambda)}dX^2 +
   Y^2(\lambda)X^{2\delta}\left( dy^2+
   k^{-1}\sin{(\sqrt{k}y)}dz^2 \right)\ ,
\ee
where $\lambda = X/T$ and $\delta = (a + f)/f$. The relation $ak = 0$ thus
takes the form $(\delta - 1)k = 0$.
The fluid velocity, $u^a$, is given by $e^{-\Psi/2}(1,0,0,0)$.

The following transformation:
\be
  X = e^{-f(x - F({\bar t}))}\ ,\qquad  \lambda = X/T = e^{f{\bar t}}\ ,
\ee
where
\be
 \frac{dF}{d{\bar t}} =
 \frac{e^{\Psi - 2f{\bar t}}}{e^\Lambda - e^{\Psi - 2f{\bar t}}}\ ,
\ee
leads to the ``homothetic'' adapted SSS line element
\be
   d{\tilde s}^2 = e^{-2fx} ds^2 = e^{-2fx}\left[
   -N^2d{\bar t}^2 + D_1{}^2 dx^2 + D_2{}^2 e^{-2ax}\left( dy^2+
   k^{-1}\sin{(\sqrt{k}y)}dz^2 \right)\right]\ ,
\ee
where
\be\eqalign{
 N^2 &=
 f^2e^{-2fF({\bar t}) + \Lambda + \Psi - 2f{\bar t}}
 (e^\Lambda - e^{\Psi - 2f{\bar t}})^{-1}\ ,\cr
 D_1^2 &=
 f^2e^{-2fF({\bar t})}(e^\Lambda - e^{\Psi - 2f{\bar t}})\ ,\quad
 D_2^2 = e^{-2f\delta F({\bar t})}Y^2\ .\cr
}\ee
The fluid velocity $v$ is given by
$v^2 = e^{-\Lambda + \Psi - 2f{\bar t}}$.

\end{document}